\documentclass[sigconf, screen, natbib=false,authorversion,nonacm]{acmart}
\usepackage[dvipsnames,svgnames]{xcolor}
\makeatletter
\ifnum\pdf@shellescape=1\relax\else\PassOptionsToPackage{draft}{minted}\fi
\makeatother

\usepackage{soul}
\usepackage{xspace}

\usepackage{listings-rust}
\usepackage{siunitx}
\usepackage[capitalise,noabbrev,nameinlink]{cleveref}

\usepackage{paralist}

\usepackage{cleveref}
\usepackage{siunitx}
\usepackage{enumitem}
\usepackage{multirow}
\usepackage{adjustbox}
\usepackage{algorithm}
\usepackage{algpseudocode}

\algnewcommand\algorithmicforeach{\textbf{for each}}
\algdef{S}[FOR]{ForEach}[1]{\algorithmicforeach\ #1\ \algorithmicdo}

\usepackage{hyphenat}
\usepackage{todonotes}[disabled=true]

\makeatletter%
\if@todonotes@disabled%
\else%
\fi
\makeatother

\newcommand\tool{\textsc{FuzzDelSol}\xspace}
\newcommand\ledgersnap{ledger snapshot\xspace}
\newcommand\blockemu{blockchain emulator\xspace}

\newcommand\noc{\num{6049}\xspace}

\newcommand\circled[1]{%
    \tikz[baseline=(char.base)]{
        \node[shape=circle,draw,inner sep=.5pt] (char) {#1};
}}

\newcommand{\tno}{{\color{red}{$\times$}}\xspace}
\newcommand{\tyes}{{\color{teal}{$\checkmark$}}\xspace}

\newcounter{challenges}
\newenvironment{Challenge}{%
  \refstepcounter{challenges}%
}{}

\crefname{challenge}{challenge}{challenges}

\newcounter{cfinding}

\crefname{cfinding}{finding}{findings}
\Crefname{cfinding}{Finding}{Findings}

\newcommand{\finding}[1]{%
	\refstepcounter{cfinding}%
	\paragraph{Finding~\thecfinding: #1}%
}

\newcommand{\settikzmark}[1]{%
  \tikz[overlay,remember picture,baseline] \coordinate (#1) at (0,0) {};}

\newcommand{\highlight}[2]{%
  \draw[YellowGreen,line width=6pt,opacity=0.2]%
    ([yshift=1.5pt]#1) -- ([yshift=1.5pt]#2);%
}

\usepackage{listings}

\definecolor{lstgreen}{rgb}{0,0.6,0}
\lstdefinestyle{plain}{%
  numbers=none,
  frame=none,
  xleftmargin=1pt,
  xrightmargin=1pt,
}

\lstdefinelanguage{Rust}{%
  morekeywords={as,break,const,continue,crate,else,enum,extern,false,
    fn,for,if,impl,in,let,loop,match,mod,move,mut,pub,ref,return,Self,
    self,static,struct,super,trait,true,type,unsafe,use,where,while,
    abstract,alignof,become,box,do,final,macro,offsetof,override,priv,
    proc,pure,sizeof,typeof,unsized,virtual,yield},
  morekeywords=[2]{isize,usize,char,bool,str,String,u8,u16,u32,u64,u128,i8,i16,i32,i64,i128,f32,f64},
  sensitive=true,
  morecomment=[l]{//},
  morecomment=[l]{///}, % change color
  morestring=[b]{"},
}%

\usepackage{minted}
\setminted{
  escapeinside=||,
  fontsize=\scriptsize,
  frame=single,
  numbers=left,
  numbersep=1ex,
}

\renewcommand{\paragraph}[1]{{\smallskip\noindent\bf{#1.}}}

\usepackage{float}
\newfloat{algorithm}{t}{lop}

\copyrightyear{2023}
\acmYear{2023}
\setcopyright{acmlicensed}
\acmConference[CCS '23] {Proceedings of the 2023 ACM SIGSAC Conference on Computer and Communications Security}{November 26--30, 2023}{Copenhagen, Denmark.}
\acmBooktitle{Proceedings of the 2023 ACM SIGSAC Conference on Computer and Communications Security (CCS '23), November 26--30, 2023, Copenhagen, Denmark}
\acmISBN{979-8-4007-0050-7/23/11}
\acmDOI{10.1145/3576915.3623178}

\usepackage[
backend=biber,
sortcites=true,
]{biblatex}
\usepackage{csquotes}
\begin{document}

\title{Fuzz on the Beach: Fuzzing Solana Smart Contracts}

\author{Sven Smolka}
\affiliation{%
 \institution{University of Duisburg-Essen}
 \country{}
}
\email{sven.smolka@uni-due.de}

\author{Jens-Rene Giesen}
\affiliation{%
 \institution{University of Duisburg-Essen}
 \country{}
}
\email{jens-rene.giesen@uni-due.de}

\author{Pascal Winkler}
\affiliation{%
 \institution{University of Duisburg-Essen}
 \country{}
}
\email{pascal.winkler@uni-due.de}

\author{Oussama Draissi}
\affiliation{%
 \institution{University of Duisburg-Essen}
 \country{}
}
\email{oussama.draissi@uni-due.de}

\author{Lucas Davi}
\affiliation{%
 \institution{University of Duisburg-Essen}
 \country{}
}
\email{lucas.davi@uni-due.de}

\author{Ghassan Karame}
\affiliation{%
 \institution{Ruhr University Bochum}
 \country{}
}
\email{ghassan@karame.org}

\author{Klaus Pohl}
\affiliation{%
 \institution{University of Duisburg-Essen}
 \country{}
}
\email{klaus.pohl@uni-due.de}

\begin{abstract}
  %!TEX root = ./main.tex

Solana has quickly emerged as a popular platform for building decentralized applications (DApps), such as marketplaces for non-fungible tokens (NFTs).
A key reason for its success are Solana's low transaction fees and high performance, which is achieved in part due to its stateless programming model. Although the literature features extensive tooling support for smart contract security, current solutions are largely tailored for the Ethereum Virtual Machine. Unfortunately, the very stateless nature of Solana's execution environment introduces novel attack patterns specific to Solana requiring a rethinking for building vulnerability analysis methods. 

In this paper, we address this gap and propose \tool, the first binary-only coverage-guided fuzzing architecture for Solana smart contracts. \tool faithfully models runtime specifics such as smart contract interactions. Moreover, since source code is not available for the large majority of Solana contracts, \tool operates on the contract's binary code. Hence, due to the lack of semantic information, we carefully extracted low-level program and state information to develop a diverse set of bug oracles covering all major bug classes in Solana. Our extensive evaluation on \noc smart contracts shows that \tool's bug oracles finds impactful vulnerabilities with a high precision and recall.
To the best of our knowledge, this is the largest evaluation of the security landscape on the Solana mainnet.
\end{abstract}

\maketitle
%!TEX root = ../main.tex

\section{Introduction}
Smart contracts are an essential part of the ecosystem in many modern blockchain platforms.
Smart contracts allow developers to implement decentralized applications (DApps) that encode business logic on the blockchain, thereby facilitating a number of use cases.
For instance, smart contracts enable the creation of non-fungible token (NFT) marketplaces.
Artists use NFT marketplaces to auction their creations.
Furthermore, established companies and sport franchises, like Nike~\cite{nike-nft}, Budweiser~\cite{budweiser-nft}, Lacoste~\cite{lacoste-nft}, and the NBA~\cite{nba-nft} use these marketplaces to sell NFT collections to fans and investors all over the world. 
 
The Solana blockchain~\cite{solana} has become a key platform in the DApps and NFT space, because of its high performance and low transaction fees.
In comparison to the more established smart contract platform Ethereum~\cite{ethereum}, Solana can execute 100--1000 times more transactions per second~\cite{Rouhani2017-bg,Lee_undated-wi} while charging a fraction of a USD cent as a fee~\cite{solana,Lee_undated-wi}.
As a result, the number of all transactions in the Solana network significantly exceeds the number of all transactions made in Ethereum by a factor of 85\footnote{On April 12, 2023, Ethereum processed \num{1933} million (\url{https://etherscan.io}) transactions, while Solana already reached \num{164839} million transactions (\url{https://solscan.io}).}.

From a smart contract perspective, the Solana platform achieves a high transaction rate because its execution layer decouples program logic from state, i.e., smart contracts cannot store any dynamic state.
This enables Solana to execute transactions operating on different data in parallel. 
However, this also introduces new attack patterns that are specific to Solana.
In fact, attacks against Solana smart contracts already caused multi-million Dollar losses~\cite{White2022-mango,White2022-solend}---the popular Wormhole attack induced losses of up to 320 million USD~\cite{Goodin2022-wormhole}.

Smart contract security research ranges across different disciplines: from formal verification~\cite{Schneidewind2020-mj} and static analysis~\cite{Torres2018-je, Mossberg2019-xp} to dynamic analysis~\cite{Ding2021-qg}, with a high focus on the Ethereum platform.
However, Solana suffers from different vulnerabilities than Ethereum, and the aforementioned techniques are not applicable to Solana due to its unique features:
In comparison to Ethereum, the Solana blockchain is stateless and smart contracts have no direct association with the state.
The stateless nature of Solana's execution environment requires stricter handling of user input.
However, vulnerabilities often come from developers not checking security-critical properties in smart contracts, like missing transaction \emph{signer checks}.
Unlike Ethereum smart contracts, which implicitly trust their state if it is not compromised.
Moreover, the stateless approach of Solana and the impact on its security model is largely unexplored.

Research on Solana security and tooling is limited: At the time of writing, VRust~\cite{Cui2022-nm} is the only existing static analysis approach that covers Solana smart contracts.
VRust incorporates detection patterns for common vulnerabilities in Solana smart contracts and was able to detect 12 vulnerabilities in popular open-source smart contracts~\cite{Cui2022-nm}.
However, VRust suffers from several limitations:
\begin{inparaenum}[1)]
  \item it strictly requires source code to conduct analyses,
  \item it suffers from a high number of false alarms, and
  \item it does not provide an analyst with enough data to (re-)construct exploit transactions.
\end{inparaenum}
In contrast, fuzzing is a technique that does not suffer from any of these limitations~\cite{Rawat2017-lj,Aschermann2019-ha,aflpp,libafl}.
The fuzzing input given to the analysis target can also usually be crafted into exploit transactions.
Smart contract fuzzing is a valuable technique that has been extensively researched with promising results.~\cite{sfuzz, echidna, rodler2023efcf}.

\paragraph{Contributions}
In this work, we propose a solution to detect bugs in Solana called \tool: the \emph{first} binary-only coverage-guided fuzzer for Solana smart contracts.
We developed a set of bug detection oracles to facilitate the detection of Solana-specific smart contract bugs, namely 
\begin{inparaenum}[1)]
  \item missing signer checks, that is, the smart contract performing critical operations without checking for signatures,
  \item missing owner checks, which allow a smart contract to use untrusted data,
  \item arbitrary cross program invocation, i.e., a smart contract calls \emph{any} other smart contract,
  \item missing key checks, which, similar to the missing owner check, enables a smart contract to use spoofed accounts as system variables, and
  \item integer bugs.
\end{inparaenum}
In addition, we also design a generic oracle for \tool to detect vulnerabilities based on lamport gains.
Lamports are the smallest denomination of Solana's native currency SOL, and 1 SOL corresponds to \num{1000000000} (one billion) lamports.
We use this oracle to detect arbitrary leaking funds and lamport-theft.

Our extensive evaluation of \tool consists of several experiments.
In our first experiment, we test \tool with a set of vulnerable smart contracts provided by the community~\cite{Neodyme2021-vw}.
We demonstrate that our approach is capable of quickly detecting Solana-specific vulnerabilities.
Compared to VRust, \tool does not report any false alarm for the dataset provided in~\cite{Neodyme2021-vw}. In addition, \tool is also able to precisely trace back the vulnerability classes.

Next, we perform a large-scale bug-finding evaluation on all Solana smart contracts present on the mainnet on March 27, 2023, which amounts to a total of \noc smart contracts.
At the time of writing, this is the largest analyzed dataset of Solana smart contracts.
\tool reports 92 bugs in these smart contracts.
We analyzed 16 reports in-depth and confirmed that only 2 were false alarms, thus demonstrating the high accuracy of \tool in detecting bugs.
\tool is the only analysis tool available that is able to analyze these contracts on the mainnet. 

Third, our performance evaluation on 16 smart contracts from well-known bug bounty programs demonstrates that \tool is able to analyze complex smart contracts.
Here, \tool's generated transactions are able to consistently find new code paths in the programs. 
In this experiment, \tool reports a true bug and only has a single false alarm, i.e., a single wrongly reported vulnerability.
We summarize our contributions as follows:

\begin{itemize}
  \item We present the \emph{first} fuzzing architecture for Solana smart contracts.
  We conceptualize \tool (\Cref{sec:design}) around the original Solana runtime to \tool faithfully model runtime specifics, such as smart contract interaction.
  Moreover, this design guarantees reproducibility and validity of transactions that \tool generates:
  Every transaction that generates a vulnerability report can be replayed, e.g., on a test network.
  \item We design and implement new bug oracles to detect Solana-specific vulnerabilities (\Cref{sub:sol-vulnerabilities}).
  Due to our design choices, \tool detects impactful bugs in smart contracts regardless of source code availability.
  \item Our extensive evaluation on \noc smart contracts shows that \tool's bug oracles find bugs with a high precision and recall.
  This is the largest evaluation of the security landscape on the Solana mainnet.
  \item \tool detects the infamous Wormhole bug.
\end{itemize}
\section{Solana's Execution Environment}
\label{sec:sol-analysis}
In the following, we provide an overview of Solana's execution environment~\cite{solana-doc, solana-rbpf}.

\paragraph{Solana Account Model}
The Solana account model decouples accounts containing non-executable raw data from accounts containing executable code.
In order to achieve this decoupling, Solana introduces an account layout consisting of the following six fields:
\begin{inparaenum}[(1)]
    \item the public key of the account,
    \item the public key of the account owner, 
    \item the executable flag, which indicates whether an account is executable (and accordingly a program),
    \item the rent epoch, which specifies a point in time when the account must pay rent to remain deployed on the blockchain,
    \item the funds that the account holds, in a unit called \emph{lamports}, and
    \item the data that the account contains.
\end{inparaenum}

Smart contracts in Solana are called \emph{programs}.
Executable on-chain programs contain extended Berkeley Packet Filter (eBPF) bytecode compiled as an executable and linkable format (ELF) shared object file in their data field and are stateless, i.e., they do not store runtime-modifiable data in their data field.
Programs manage runtime-modifiable data in non-executable accounts whose owner field contains the program's public key.
Once a program is the owner of an account, only that program is able to modify the account's data as well as deduct lamports from the account. 

Solana distinguishes between two types of programs: \emph{i)} \emph{native programs} and \emph{ii)} \emph{on-chain programs}. 
Native programs are implemented in the Solana runtime and are not deployed on the blockchain.
These programs typically perform tasks such as allocating new accounts on the blockchain or deploying on-chain programs on the blockchain.
On-chain programs are written in C, C++, or Rust, compiled in eBPF, and deployed on the blockchain using a native program.
Both native and on-chain programs are marked as executable.
But only on-chain programs contain eBPF bytecode in their data field. 
Moreover, on-chain programs are only capable of modifying the data and lamports of non-executable accounts at runtime.
However, some native programs are also capable of modifying the remaining four fields.

\paragraph{Solana Transactions}
Transactions consist of, but are not limited to, 
\begin{inparaenum}[(1)]
    \item a list of signatures of accounts that signed the transaction
    \item 
    a recent blockhash used to determine if the transaction is too old,
    \item a sorted list of accounts that can be used in the instructions of the transaction
    \item and a list of \emph{instructions}. 
\end{inparaenum}
An \emph{instruction} is responsible for invoking a native or an on-chain program. 
Here, a native program is executed directly in the Solana runtime, while an on-chain program is executed in Solana's eBPF VM using the eBPF bytecode stored in the program's data field.
Instructions consist of three elements: \emph{i)} the public key of the program called by the instruction (also referred to as the \emph{program id}), \emph{ii)} a list of accounts passed to the program, which must be a subset of the sorted list of accounts in the transaction, and \emph{iii)} instruction data representing arbitrary data. 
A called program accesses these three elements during execution.

\paragraph{Cluster Information}
Solana programs receive cluster information by calling functions of \emph{sysvar} accounts.
For instance, the amount of lamports to pay for allocating one data byte for an account.
According to the Solana programming model, \emph{sysvar} accounts must also be passed as input to the program by including them in the instruction's account list.

\paragraph{Program-Derived Addresses}
In addition to addresses that are a pair of public and private keys, Solana introduces Program-Derived Addresses (PDAs).
PDAs are addresses that are not located on the ed25519 curve, do not have a corresponding private key, and are associated with a program. 
A PDA receives its association with a program during its derivation, where the PDA is deterministically derived based on the program's public key and a set of optional seeds.
In order to ensure that a PDA does not lie on the ed25519 curve, a bump byte is iteratively determined that will uniquely \enquote{bump} the PDA out of the ed25519 curve.

\begin{figure*}
\centering
\includegraphics[width=0.8\linewidth]{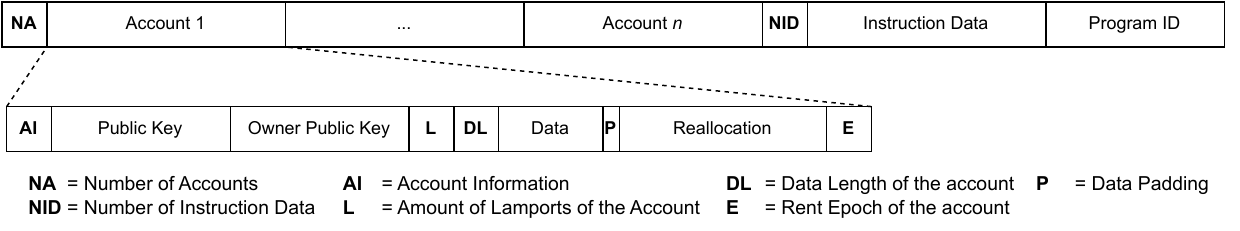}
\caption{Serialization of an instruction as input of the eBPF VM}
\label{fig:program_input_layout}
\end{figure*}

\paragraph{Cross-Program Invocation}
\label{par:cpi}
Solana programs can call other programs using Cross-Program Invocation (CPI).
Here, the caller invokes the callee with a self-created instruction that contains at least the same privileges as the instruction which invoked the caller, i.e., accounts that signed the transaction.
However, the Solana runtime allows the caller to delegate additional privileges to the callee using PDAs.
These additional privileges are restricted to accounts whose public key is a PDA associated with the caller, i.e., derived from the caller's public key along with a set of optional seeds.
Thus, when invoking a program using CPI, the calling program can use PDAs to \enquote{sign} accounts in the CPI instruction.

\paragraph{eBPF VM}
Solana's eBPF VM~\cite{solana-rbpf} can execute on-chain programs in both \emph{Just-in-time} compilation and \emph{Interpreter} mode.
The Solana runtime defines a number of environmental restrictions when executing on-chain programs in the eBPF VM. 
By default, the eBPF VM limits the resource consumption of an instruction to a maximum of a certain number of \emph{compute units}.
The Solana runtime accumulates compute units for all instructions within a transaction, with certain runtime operations such as function or system calls consuming a specific number of compute units.
In addition, the Solana runtime enables the eBPF VM to allocate up to 64 stack frames and reach a maximum depth of four Cross-Program Invocations.
Lastly, it prohibits reentrancy during CPI.

When executing an instruction in the eBPF VM, the instruction is serialized and passed to the VM as input for the program.
The program input starts at a fixed address of the VM memory layout. 
\cref{fig:program_input_layout} depicts the layout of the serialized instruction, which is divided into three parts: accounts, instruction data, and program id which corresponds to the input of the program. 
The first 8 bytes of each account contain information about whether the account has signed the transaction, is read-only, or is executable.
Followed by the public key of the account and the account's owner, the amount of lamports, the data length, and other information.
%!TEX root = ../main.tex
\section{Solana Program Security and Challenges}

There are different types of vulnerabilities in Solana programs.
Attackers exploiting these vulnerabilities may compromise accounts managed by vulnerable programs, i.e., by stealing an account's funds or manipulating an account's data. 
The attacker typically accomplishes this by crafting and executing sequences of instructions that allow the attacker to maliciously select the order of accounts and the instruction data in each instruction. 
As a result, an attacker gains access to accounts owned by or authorized to the vulnerable program, and can steal lamports or violate the accounts' data integrity.

\subsection{Solana Program Vulnerabilities}
\label{sub:sol-vulnerabilities}
In the following, we describe five key vulnerabilities~\cite{Neodyme2021-vw, Cui2022-nm, Goodin2022-wormhole} of Solana programs, which we all address in this paper. 

\paragraph{Missing Signer Check}
A \emph{missing signer check} (MSC) vulnerability exists when a program does not verify that an account that should have signed the transaction according to the business logic actually signed the transaction.
This allows an attacker to gain unauthorized access to program behavior. 

\paragraph{Missing Owner Check}
A \emph{missing owner check} (MOC) vulnerability exists when a program reads and processes data from an account that, according to the business logic, should be owned by the program without verifying it is actually owned by the program.
Thus, an attacker could create his own account with data and pass it to the program as input. 

\paragraph{Arbitrary CPI}
An \emph{arbitrary CPI} (ACPI) vulnerability exists when a program does not verify the program id of the invoked program during CPI.
Hence, an attacker can deploy a malicious program on the blockchain and pass it as input to the program possessing the arbitrary CPI vulnerability. 
As a result, the program possessing the vulnerability invokes the malicious program using CPI, giving the attacker control over the execution.
This is especially critical if the program grants additional privileges to the invoked program by signing PDAs. 

\paragraph{Missing Key Check} 
A \emph{missing key check} (MKC) vulnerability exists when a program expects a specific account and processes its data without verifying that the passed account is actually the expected specific account.
This vulnerability resulted in a loss of up to 320 million USD in the Wormhole program ~\cite{Goodin2022-wormhole}.

\paragraph{Integer Bugs}
An \emph{integer bug} (IB) exists when values underflow or overflow in arithmetic operations. 
This vulnerability can be exploited by an attacker when a program transfers lamports from account $a$ to account $b$. 
Here, the attacker selects the value to be transferred so high that the lamports field of account $b$ overflows, while the lamports field of account $a$ underflows.
The Solana runtime allows this transfer because the total amount of lamports remains the same before and after executing the instruction.

\subsection{Solana Security Analysis}
\label{sec:solana-security-analysis}

In contrast to the available tools and approaches for Ethereum smart contract security analysis, only very few approaches and tools support Solana program security analysis.
The existing ecosystem merely consists of bytecode lifters~\cite{bn-ebpf-solana,ghidra-ebpf} that are capable to lift the eBPF bytecode of Solana programs to another, tool-specific intermediate language (IL), such as Ghidra's~\cite{ghidra} or Binary Ninja's~\cite{binja} IL.
These plugins help in reverse engineering, but do not conduct any sophisticated security analysis on their own.

Currently, VRust~\cite{Cui2022-nm} is the only existing static analysis approach that covers Solana programs with a focus on security.
VRust covers a wide range of vulnerability patterns for common vulnerabilities in Solana smart contracts.
As a result, VRust was able to detect 12 vulnerabilities in popular open-source smart contracts~\cite{Cui2022-nm}.
VRust uses the Rust compiler to analyze Solana programs on the Rust Mid-Level Intermediate Representation (MIR)~\cite{rust-mir}. 
The MIR is a typed language that has complete information about the memory layout and types of a variable's data structure.

The Rust compiler uses MIR as a compiler-internal representation that models control-flows and data-flows for further optimization of the ownership and borrow checking rules, hence it is designed to model control-flows and data-flows precisely.
In general, it is possible to map MIR-level objects backwards to higher-level representations, which also provides access to high-level information like variable names.
As a result, today's decompiling and lifting approaches~\cite{bn-ebpf-solana,ghidra-ebpf} cannot achieve the same precision for data-flows as MIR.
Further, custom analysis passes can be implemented as a visitor on top of the MIR-level control flow graph for any given Solana program written in the Rust programming language.
However, relying on the MIR for vulnerability analysis strictly requires source code, because the MIR can only be generated by the Rust compiler from the higher-level representations.

Solana programs can be written in C or Rust, and it is possible to differentiate between programs written in either language by checking which system calls a program uses.
As this can be done on bytecode level, we count the occurrences of the C vs. Rust variants of the \emph{sol\_invoke\_signed} system call.
We find that 97\% of Solana programs are written in the Rust programming language, whereas the remaining programs use the C programming language.
Thus, VRust cannot analyze these programs.
The \emph{Solana security.txt}~\cite{solana-security-txt} feature allows program authors and developers to provide information on where to find the source code of a program and whom to contact for security issues.
We use this information to find out about source code availability of Solana programs.
Our findings show that less than 2\% of Solana programs come with source code.
As a result, VRust cannot be applied to a vast majority of Solana programs.

\subsection{Challenges of Fuzzing Solana Programs}\label{sec:challenges}
A popular approach to uncover bugs is coverage-guided fuzzing~\cite{afl, aflpp, nyx, libafl}.
This technique mutates the inputs based on instruction-coverage data, or feedback information, collected during the target's execution, to uncover new paths in the application.
Fuzzing Solana contracts is yet to be explored.

\paragraph{Ethereum Smart Contract Fuzzing}
Fuzzing Ethereum smart contracts is a heavily researched area~\cite{rodler2023efcf, sfuzz, harvey, fuzz-symex, contractfuzzer, confuzzius}.
Ethereum fuzzers like EFCF~\cite{rodler2023efcf} model interaction between smart contracts to detect complex and hard-to-find Ethereum bugs, e.g., compositional reentrancy bugs.
However, these approaches are not feasible or applicable to Solana, because of its unique programming model (cf. \Cref{sec:sol-analysis}).
Modelling the interaction between Ethereum smart contracts does not require in-depth information about the type of other smart contracts.
In order to model the interaction between programs and accounts in Solana, it is necessary to know the domain of a program.
For instance, the program in \Cref{lst:msc} interacts with three different types of accounts: a wallet, a vault, and an authority.
Ethereum encapsulates the state within the smart contract, i.e., an Ethereum fuzzer~\cite{contractfuzzer} does not need to consider the different types of data to detect bugs.
Nevertheless, to detect more complex bugs, like delegated re-entrancy bugs, Ethereum fuzzers~\cite{rodler2023efcf} must understand the semantics of the respective contracts.
In contrast, it is hard to detect which type an account assumes in a Solana program.
Therefore, modelling the blockchain state and the content of accounts is essential to detect real, reproducible and impactful bugs in smart contracts.
Thus, in order to faithfully fuzz Solana programs, a Solana fuzzer has to solve the following challenges:

\begin{Challenge}
\paragraph{Challenge 1: Modeling Ledger Snapshots} 
As Solana requires programs to store data in external, non-executable accounts, it is necessary to model a valid \ledgersnap which consists of multiple accounts.
Moreover, programs must be able to change this snapshot across multiple transactions as well as operate on the changed snapshots to be able to execute business logic which depends on a specific state of the \ledgersnap.
\label[challenge]{c1}
\end{Challenge}

\begin{Challenge}
\paragraph{Challenge 2: Reproducibility of Transactions} 
To identify vulnerabilities exploitable in the real production blockchain, it is necessary to generate transactions that are reproducible in the production blockchain.
\label[challenge]{c2}
\end{Challenge}

\begin{Challenge}
\paragraph{Challenge 3: Cluster Information} 
Several Solana programs require cluster information at runtime, such as the amount of lamports to allocate a byte of data. 
Solana programs receive this cluster information using functions provided by \emph{sysvar} accounts, which must be passed to the program as input.
Hence, it is necessary to ensure that Solana programs receive sysvar accounts as input and are able to call functions of these sysvar accounts. 
\label[challenge]{c3}
\end{Challenge}

\begin{Challenge}
\paragraph{Challenge 4: Solana-specific Vulnerabilities} 
Since Solana programs comprise specific vulnerabilities resulting from the Solana programming model, it is necessary to develop detection mechanisms to detect these Solana-specific vulnerabilities. 
\label[challenge]{c4}
\end{Challenge}

\begin{Challenge}
\paragraph{Challenge 5: Program semantics need to be retrieved} 
Solana programs manage accounts associated with the program using PDAs with a program-specific seed structure. 
Since programs verify whether the public key of a passed account corresponds to the program-specific PDA seed structure, it is necessary to \emph{i)} determine the program-specific PDA seed structure and \emph{ii)} to pass accounts to the program, whose public key is derived from the program-specific PDA seed structure.
\label[challenge]{c5}
\end{Challenge}

\begin{Challenge}
\paragraph{Challenge 6: Faithful CPI} 
Solana programs are able to invoke other programs at runtime using CPI. 
Hence, the runtime environment must ensure that programs can use the CPI mechanism to invoke both on-chain programs---in a separate eBPF VM---and native programs in the Solana runtime. 
\label[challenge]{c6}
\end{Challenge}

\section{Overview of \NoCaseChange{\tool}}
\label{sec:design}

In this section, we introduce the design of \tool and its main components.
Furthermore, we describe how our design tackles the challenges mentioned in \Cref{sec:challenges}.

\paragraph{Intended Use of \tool}
FuzzDelSol aims to explore the prevalence of vulnerabilities in Solana programs. 
Although similar bytecode-based security studies have been conducted for other blockchain platforms, such as Ethereum~\cite{ethbmc,Schneidewind2020-mj}, or other application domains like Android~\cite{enck2011study}, there does not yet exist any comprehensive study about the security of Solana programs. 
Hence, for the first time, we aim to raise awareness for Solana program security with \tool. 
For 98\% of Solana programs, no source code is available (see \Cref{sec:solana-security-analysis}).
Hence, we argue that source code-based analysis techniques like VRust~\cite{Cui2022-nm} are not applicable to analyze the vast majority of Solana programs. 
Moreover, \tool can be used to find vulnerabilities with the intention of forming a better understanding of the vulnerability types.
This is necessary to develop appropriate countermeasures for Solana-specific vulnerabilities.
Further, Solana program developers may use \tool to vet closed-source third-party programs interacting with their own programs.
The same applies to users of Solana programs: \tool helps in ensuring that closed-source programs a user wants to invest in are secure, before investing funds. 

\begin{figure*}
\centering
\includegraphics[width=.8\linewidth, keepaspectratio]{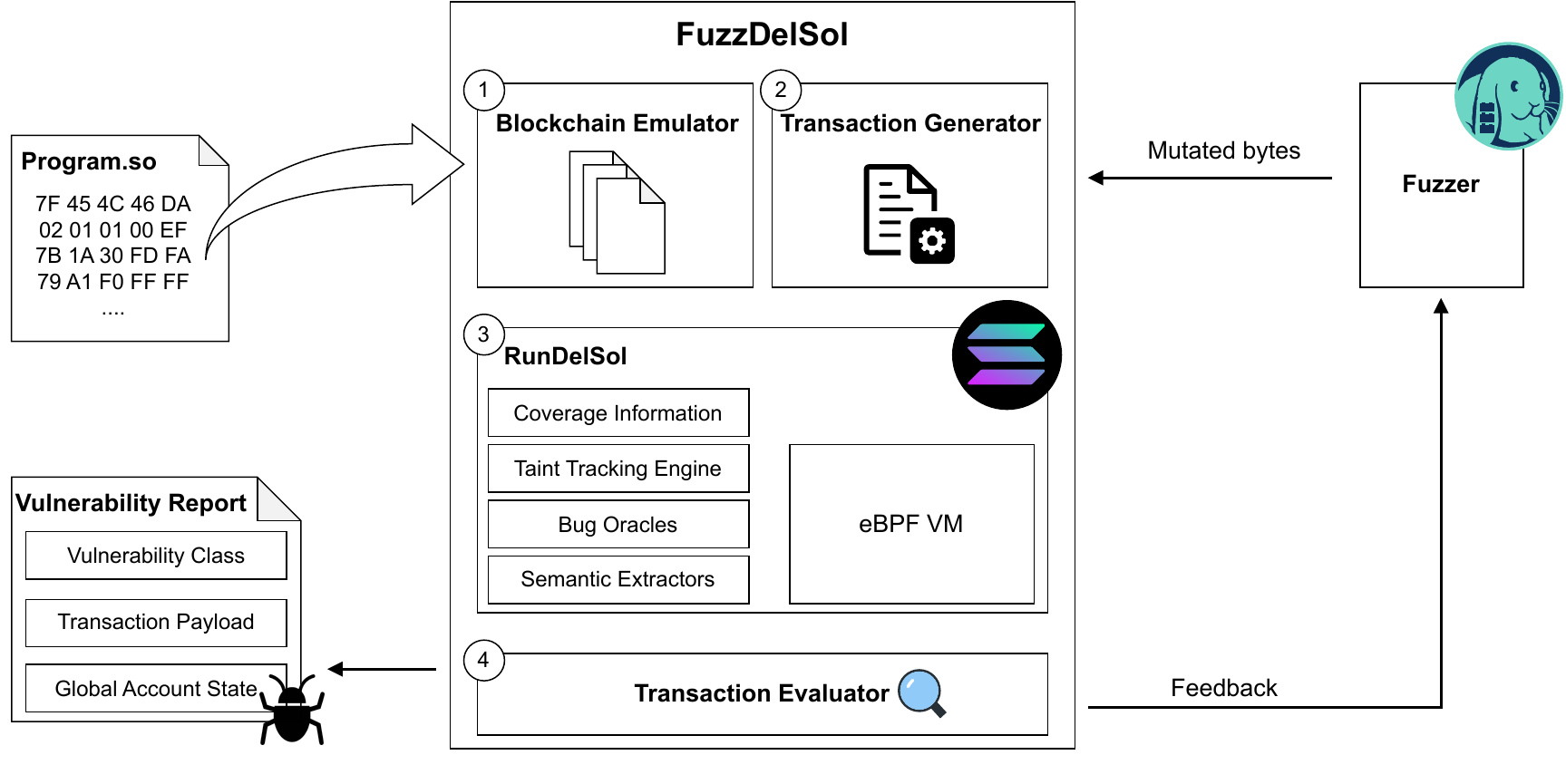}
\caption{\tool Design}
\label{fig:architecture}
\end{figure*}

\paragraph{Overview}
Our high-level architecture is shown in \Cref{fig:architecture}. 
The main idea of \tool is to \circled{1} create a valid blockchain snapshot using the \emph{blockchain emulator}; comprising the Solana program to analyze, an \emph{attacker} account and additional accounts, e.g., user and non-executable data accounts, or executable programs.
The next component is the \emph{transaction generator} \circled{2}, which receives random and mutated bytes from a fuzzer to generate valid transactions. 
\tool executes these transactions in an instrumented Solana runtime \circled{3} called \emph{RunDelSol}.
In particular, we extended the original Solana runtime with patches to detect Solana-specific vulnerabilities (cf.~\Cref{sub:sol-vulnerabilities}) induced by the generated transactions.
Finally, the \emph{transaction evaluator} \circled{4} analyzes the aftermath of the transactions, and extracts valuable insights for following fuzzing iterations.
However, if the transactions signal an erroneous \ledgersnap, \tool generates a vulnerability report with information to reproduce this \ledgersnap. 

\paragraph{Blockchain Emulator (\Cref{sec:blockchain-init})}
A challenge for fuzzing Solana programs is that program execution is largely dependent on the \ledgersnap. 
Thus, we developed a component called \emph{\blockemu}~\circled{1} to prepare the snapshot of the ledger available for analysis.
The modeled \ledgersnap contains the program being fuzzed as well as additional accounts that are relevant to the execution context.
Moreover, this component provides all public keys that can be passed to the program as input.
Operating on a valid \ledgersnap allows a program, for example, to manage and modify lamports and account data at runtime across multiple transactions. 
\tool uses the \blockemu to model a \ledgersnap for programs to operate on during program execution, thereby addressing~\Cref{c1}.
In addition, the \blockemu---along with RunDelSol---also enables \tool to address \Cref{c3}, since the public keys provided by the \blockemu include those of the sysvar accounts used by the program to retrieve cluster information. 
Furthermore, the \blockemu incorporates an account generator that creates attacker-controlled accounts containing malicious data to trigger Solana-specific vulnerabilities.
Therefore, the \blockemu supports addressing \Cref{c4}.
Finally, the \blockemu uses PDA seed structures obtained from the transaction evaluator to derive valid PDAs, which assists in tackling \Cref{c5}.

\paragraph{Transaction Generator (\Cref{sec:transaction-gen})}
For each fuzzing iteration, \tool \emph{mimics} real Solana transactions to find reproducible bugs.
However, Solana transactions contain structured data.
Thus, \tool incorporates a \emph{transaction generator}~\circled{2}, which \emph{transforms} the randomly generated bytes from a fuzzer into valid Solana transactions.
As the transaction generator produces valid and reproducible transactions, \tool covers \Cref{c2}.

\paragraph{RunDelSol (\Cref{sec:rundelsol})}
Effective fuzzing requires coverage feedback~\cite{afl, aflpp} to guide the generation of test inputs. 
Source code is commonly used to instrument a program for achieving accurate results in this feedback mechanism.
However, given the absence of source code for the large majority of Solana programs, we cannot rely on the source code.
Therefore, instead of instrumenting programs, \tool implements a specialized Solana runtime environment~\circled{3}, called \emph{RunDelSol}.
This environment---besides instantiating and executing Solana programs on the previously generated \ledgersnap---includes instrumentation to measure coverage.
Furthermore, it allows programs to invoke functions of passed sysvar accounts to retrieve \emph{cluster information} (addressing \Cref{c3} along with the \blockemu).   

Moreover, RunDelSol features a \emph{taint tracking engine} to trace the data-flow during the execution of a program.
This enables us to implement Solana program vulnerability detectors or \emph{bug oracles}, as well as Solana-specific runtime information extractors which extract PDA seed structures.
Each oracle aims to detect a potential vulnerability without source code information and uses the taint tracking engine differently, and therefore tackling \Cref{c4}. 
Moreover, the Solana-specific runtime information extractors of RunDelSol (along with the transaction evaluator, see below) allow \tool to address  \Cref{c5} by extracting PDA seed structures.

Moreover, RunDelSol can invoke multiple programs using CPI and run them in separate eBPF VMs for on-chain programs.
In the case of native programs, RunDelSol executes them in the Solana runtime.
Otherwise, for invoking on-chain programs, RunDelSol traces their data-flow using the taint tracking engine independently to the callee program.
This allows the oracle to detect vulnerabilities across multiple CPI invocations.
Enabling programs to call other programs using CPI and analyzing their interaction allows \tool to solve \Cref{c6}.

\paragraph{Transaction Evaluator (\Cref{sec:transaction-eval})}
Transactions can impact the state of the blockchain.
For correct adjustment of the \ledgersnap and preparation of the next fuzzing iteration, the \emph{transaction evaluator} \circled{4} extracts relevant information from RunDelSol after the execution of transactions, including PDA seed structures, eBPF VM signals, and feedback information for the fuzzer, e.g., coverage.
Next, the transaction evaluator forwards the information to the fuzzer and the \blockemu and decides whether it should re-generate the \ledgersnap for subsequent fuzzing iterations, taking into account the newly received semantic program information. 
As a result, the transaction evaluator, together with RunDelSol, allows addressing \Cref{c5}. 

\section{\NoCaseChange{\tool} Internals}
\label{sec:tool}
In this section, we detail the implementation of \tool.
\tool uses the state-of-the-art Libafl~\cite{libafl} fuzzer.
Libafl's design allows the \tool to include its own feedback mechanism in the fuzzing mutation.
The bytes generated by Libafl are based on feedback obtained from previous fuzzing iterations, which helps uncover new paths and overcome barriers, e.g., such as public key comparisons, in the Solana program.

We describe how each of the four components depicted in \Cref{fig:architecture} is used to detect bugs.
Moreover, we leverage an exemplary program that suffers from two impactful bugs to describe \tool for ease of presentation.
\Cref{lst:msc} shows the code of this example program.
We chose this example because it showcases two of the more popular bugs in Solana.
This function allows a user to withdraw funds from a \emph{vault} managed by the program. 
\Cref{line:check2} in the function checks if the \emph{authority} account is authorized to withdraw funds from account \emph{vault}.
In contrast to Ethereum, Solana programs need a separate wallet account to store data.
In \Cref{line:check3}, the function also checks whether the \emph{vault} provided as an input is associated with the \emph{wallet\_info} account.
Finally, if account \emph{from} contains sufficient lamports, the transfer proceeds.
However, this program suffers from a \emph{missing signer check}: an attacker can provide account \emph{authority} without the actual owner of \emph{authority} knowing about this.
Hence, the attacker can transfer funds on behalf ob the actual \emph{authority} without proper authorization.
This vulnerability can be fixed by adding a signer check to the program, which is shown in~\Cref{line:signercheck}.
Furthermore, the program lacks an owner check to verify the integrity of the information stored in the \emph{wallet\_info} account.
In this case, an attacker can provide a forged \emph{wallet\_info} account with fake data that refers to any \emph{vault} managed by the program while supplying his own public key as the \emph{authority}.
This means that, even if the code includes the signer check in \Cref{line:signercheck}, an attacker could drain any \emph{vault} associated to the program.
The check in \Cref{line:ownercheck} prevents this, as it verifies that the information in \emph{wallet\_info} is trusted.
Therefore, even if an attacker passes a fake account to the program, the program recognizes that this account does not contain the program's \emph{program\_id} as owner.
Thus, this check mitigates illegal lamport withdrawal.
Note that \tool is detecting these bugs without access to the program's source code. 

\subsection{Blockchain Emulator}\label{sec:blockchain-init}
The \blockemu allows \tool to model a valid \ledgersnap.
When creating a \ledgersnap, the \blockemu deploys several accounts: 
\begin{inparaenum}[(1)]
    \item a \emph{user} and \emph{attacker} wallet account, which are used by the oracles to determine whether a transaction triggered a potential security vulnerability
    \item on-chain programs such as the program to fuzz or the \emph{SPL Token program}
    \item \emph{sysvar} accounts, enabling programs to use their functions to receive \emph{cluster information}
    \item attacker-controlled accounts, which attempt to exploit potential missing owner check vulnerabilities.
\end{inparaenum}

%!TEX root = ../main.tex
\begin{figure}[t]
\begin{lstlisting}[language=rust, breaklines=true, escapechar = !,caption ={Solana program to withdraw funds from a vault. This program contains a missing signer and owner check. We mark additional checks with {\color{OliveGreen}+} and green highlighting.}, captionpos = b, label = {lst:msc}]
fn withdraw(program_id: Pubkey, accounts: [AccountInfo], amount: u64) -> ProgramResult {

    // The accounts from the transaction are attacker-controlled,
    // including wallet_info and wallet
    let wallet_info = accounts[0];
    let wallet = deserialize(wallet_info.data);
    
    let vault = accounts[1];
    let authority = accounts[2];
    
    // An attacker can forge the authority and vault fields 
    // of the wallet account, to pass both of these checks
    !\label{line:check2}!assert(wallet.authority == authority.key);
    !\label{line:check3}!assert(wallet.vault == vault.key);
    
!\settikzmark{patch1s}\color{OliveGreen}{+}!   // FIX: check owner of account from!\settikzmark{patch1e}!
!\label{line:ownercheck}\settikzmark{patch2s}\color{OliveGreen}{+}!   assert(wallet_info.owner == program_id);!\settikzmark{patch2e}!

!\settikzmark{patch3s}\color{OliveGreen}{+}!   // FIX: the following line adds the required signer check!\settikzmark{patch3e}!
!\label{line:signercheck}\settikzmark{patch4s}\color{OliveGreen}{+}!   assert(authority.is_signer); !\settikzmark{patch4e}!

    // check for sufficient funds
    if amount > from.lamports {
        raise InsufficientFundsException;
    }
    
    // transfer lamports
    !\label{line:vlamports}!vault.lamports -= amount;
    authority.lamports += amount;

    // XXX: Missing signer and owner check vulnerability
    // Funds can be transfered to and from unauthorized accounts
    Ok(())
}
\end{lstlisting}
\begin{tikzpicture}[remember picture, overlay]
    \highlight{patch1s}{patch1e}
    \highlight{patch2s}{patch2e}
    \highlight{patch3s}{patch3e}
    \highlight{patch4s}{patch4e}
  \end{tikzpicture}
%   \vspace{-6ex}
\end{figure}

The \blockemu provides all these accounts and their public keys to the transaction generator for generating transactions.
The example in \Cref{lst:msc} requires three accounts: the \emph{wallet\_info}, \emph{vault} and \emph{authority} accounts with their respective data.
If these accounts are not available, the transaction fails.
\tool generates all three accounts and deploys them on the blockchain.
The \blockemu additionally provides the \emph{user}'s and \emph{attacker}'s private keys, as well as the public keys of Solana's native programs---which do not need to be deployed on the blockchain as they are integrated in the Solana runtime.
Finally, it also implements a \emph{PDA generator} and an \emph{attacker-controlled accounts generator}, which receive information about the runtime from the transaction evaluator.

\paragraph{PDA Generator}
The PDA generator derives PDAs based on the information provided by the PDA seed structures. 
The seed structures are extracted by the \emph{PDA seed structure extractor} (cf. \Cref{sec:program-semantic-extractors}) and received by the transaction evaluator.
The generator creates user-related and attacker-related PDAs by inserting the user's and the attacker's public key at the precise seed positions at which the program expects a public key. 
Furthermore, the generator inserts statically extracted seeds in places where the PDA seed structure extractor could not determine the origin of the seed.
In the case that no seed exists in a PDA seed structure that originates from a public key, the PDA generator derives a single PDA without reference to the user or attacker.
Finally, the PDA generator provides each derived PDA to the transaction generator for generating transactions.

\paragraph{Attacker-controlled Accounts Generator}
This generator uses information about the location of public keys from the account data---extracted by the \emph{account data structure extractor} (cf. \Cref{sec:program-semantic-extractors}) and received by the transaction evaluator---to create attacker-controlled accounts populated with malicious data and deploys them on the \ledgersnap. 
Here, the generator populates the attacker-controlled account data with public keys of user-related accounts and the public key of the attacker at positions at which the program expects public keys.
Hence, an extracted account structure must contain at least two public keys for the generator to generate attacker-controlled accounts.
By generating attacker-controlled accounts containing malicious data, \tool aims to detect potential missing owner check vulnerabilities.
In \Cref{lst:msc}, this applies to the wallet account \emph{wallet\_info} with an account \emph{authority} by comparing \emph{wallet\_info's} data with the public key of \emph{authority}. 
Since the program does not contain an owner check, \tool treats the \emph{wallet\_info} account as attacker controlled. 
Lastly, the attacker-controlled accounts generator provides the public key of each generated attacker-controlled account to the transaction generator.

\subsection{Transaction Generator}\label{sec:transaction-gen}

\begin{figure*}[t]
\centering
\includegraphics[width=0.7\linewidth]{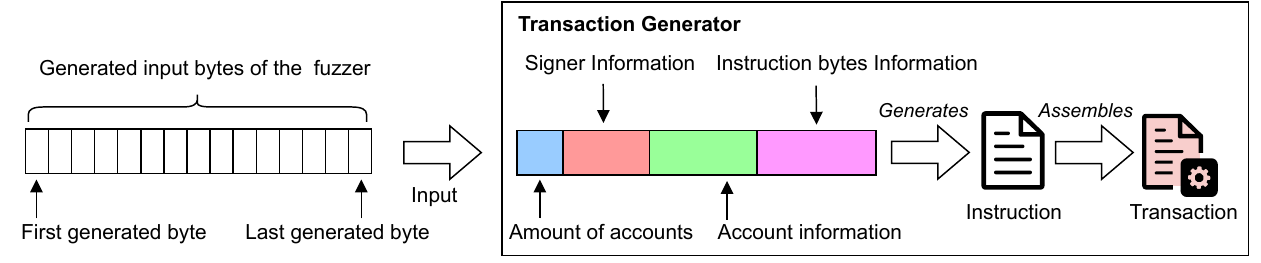}
\caption{Process to transform randomly generated bytes into a valid transaction}
\label{fig:instruction_generation_structure}
\end{figure*}
The transaction generator allows \tool to create valid and reproducible Solana transactions, effectively simulating real transactions. 
This component receives generated bytes from the fuzzer, and blockchain information from the \blockemu.
This includes information regarding the public keys of the selectable accounts, the last blockhash, and accounts capable of signing a transaction

As each transaction consists of at least one instruction, the generator first generates instructions from the received bytes, and then includes the remaining elements of a transaction, including the signature list, blockhash, and sorted account list (see \cref{sec:sol-analysis}). 

\cref{fig:instruction_generation_structure} shows the process and pattern the transaction generator applies to transform the generated bytes of the fuzzer into a valid Solana instruction.
The pattern can be mainly divided into four steps: \emph{i)} number of accounts, \emph{ii)} information about the account signing the transaction containing the instruction, \emph{iii)} information about the accounts contained in the instruction, \emph{iv)} and information about the instruction data contained in the instruction.
Regarding \Cref{lst:msc}, in step \emph{iii)}, the \blockemu extracts the accounts used in the instruction, and deploys accounts for \emph{wallet\_info}, \emph{vault}, and \emph{authority}.

For example, the number of accounts generated for the transaction is defined by a one-byte field.
This field  defines how many accounts the generator should insert in the instruction, considering the maximum number of selectable accounts provided by the \blockemu.
Similarly, the remaining bytes provided by the fuzzer are structured to create valid transactions and accounts.
Here, FuzzDelSol takes dependencies between these fields into account:
For instance, the number of generated accounts directly affects the account and signer information fields.
The instruction bytes information field uses the raw bytes generated by the fuzzer as instruction data.
After transforming the bytes into a valid instruction, with all the required information, the transaction generator creates an associated transaction for that instruction.

\subsection{RunDelSol}\label{sec:rundelsol}
RunDelSol executes the previously generated transactions in the Solana runtime environment. 
Since RunDelSol uses the real Solana environment, \tool can provide cluster information to the executed Solana program to fuzz and allow invoking other programs deployed on the blockchain (on-chain programs) and integrated in the runtime (native programs) using CPI.
Moreover, RunDelSol instantiates the Solana program to fuzz in an instrumented eBPF VM in Interpreter mode and executes it based on the \ledgersnap generated by the \blockemu. 
RunDelSol includes an instrumented eBPF VM, which extends the execution environment with \emph{coverage information}, data-flow tracing with \emph{taint tracking}, the implementation of six \emph{bug oracles} to detect potential vulnerabilities, and the extraction of \emph{Solana-specific program semantics}. 

\subsubsection{Coverage Information}
The instrumented eBPF VM of RunDelSol extracts coverage information during program execution. 
Here, RunDelSol examines each transition in the control flow graph, i.e., each \texttt{JMP}, \texttt{CALL}, and \texttt{RET} eBPF-instruction, and computes an index for that transition as follows: 
Let \emph{src} be the program counter of the eBPF \texttt{JMP}, \texttt{CALL} or \texttt{RET} instruction and \emph{dst} be the program counter of the target instruction of the transition, then the index of the transition is \( i \leftarrow (\emph{src} + \emph{dst}) \mod \emph{s}\), where \emph{s} is the size of the \emph{coverage array}.
The indices of the \emph{coverage array} provide information about whether the program executed this transition.
After the instrumented eBPF VM terminates, RunDelSol forwards the received coverage information of the executed instruction to the transaction evaluator. 

\subsubsection{Taint Tracking Engine}
\label{par:taint_tracking_engine}
The taint tracking engine provides fine-grained tracing of data-flows during program execution.
RunDelSol instruments memory addresses and register indices used by \texttt{LOAD}, \texttt{STORE} and \texttt{MOV} instructions.
The Data-flow tracing is enabled for the following three events. 
First, \tool traces data of accounts located on the \emph{program input} (cf. \cref{fig:program_input_layout}) and account public keys read by the program.
In \Cref{lst:msc} \tool traces the data of account \emph{wallet\_info}, because its data is read in \Cref{line:check2,line:check3}.
\tool also traces the public keys of the \emph{vault} and \emph{authority} accounts for the same reason.
Second, the taint tracking engine starts tainting memory addresses where the program stores the return values of Solana-specific syscalls for deriving PDAs. 
Tainting these memory addresses is crucial, as the program can use them in its execution instead of the public keys of the accounts located on the program input. 
Third, we taint register indices where the program stores an overflow or underflow value as a result of an arithmetic eBPF-instruction.

\begin{figure}[t]
\centering
\includegraphics[width=\columnwidth]{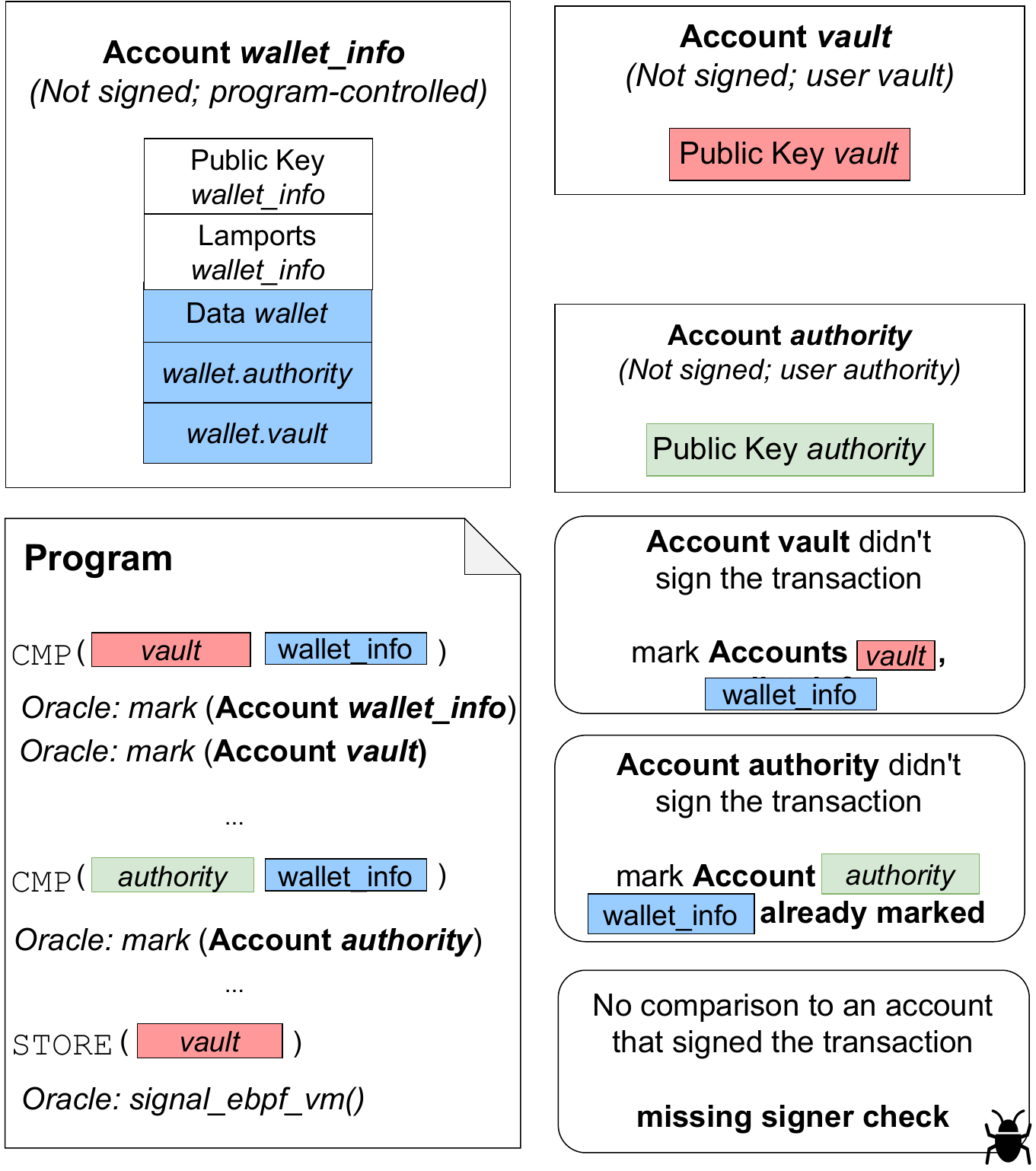}
\caption{Interaction of the taint tracking engine and the missing signer check oracle}
\label{fig:missing_signer_check_oracle_combination_taint_tracking_engine}
\end{figure}

\subsubsection{Bug Oracles}\label{sub:oracle}
RunDelSol implements six bug oracles in the instrumented eBPF VM, to find the following vulnerabilities:
\emph{i)} missing signer check, \emph{ii)} missing owner check, \emph{iii)} arbitrary cross program invocation, \emph{iv)} missing key check, \emph{v)} integer bugs, \emph{vi)}, and a lamports-based bug oracle to detect lamport-theft.
These oracles investigate eBPF-instructions, leverage the taint tracking engine to trace data-flows between eBPF-instructions, and signal the instrumented eBPF VM if a transaction triggers a potential vulnerability. 
When signaling the eBPF VM, each oracle specifies the reason for the crash and the program counter at which the crash occurs.
In the following, we will describe how these oracles work.

\paragraph{Missing Signer Check Oracle}
\Cref{lst:msc} suffers from a missing signer check.
\tool traces and analyzes the data-flow of this program to detect this vulnerability.
\Cref{fig:missing_signer_check_oracle_combination_taint_tracking_engine} describes this process:
The program reads data from the \emph{wallet\_info}, and public keys from the \emph{vault} and \emph{authority} accounts.
\Cref{algorithm:missing_signer_check} details the bug-detecting process for a missing signer check.
The oracle keeps track of mainly two things: accounts that the oracle marks as \emph{vulnerable}, and accounts that are marked as \emph{not vulnerable}.
Further, the oracle marks accounts as vulnerable if their public key or data are compared with another account without any of the accounts having signed the transaction (see Line~\ref{missing_signer_check_13}).
On the other hand, the oracle marks accounts as \emph{not vulnerable} if a comparison against an account that signed the transaction happens (see Line~\ref{missing_signer_check_16}).
The oracle signals the VM if the program writes to the lamports or data fields of an account that is marked as \emph{vulnerable} (see Line \ref{missing_signer_check_19} to \ref{missing_signer_check_21}).

Regarding the example in \Cref{lst:msc}, because of the checks in \Cref{line:check2,line:check3}, \tool marks the \emph{wallet\_info} account as \emph{vulnerable}.
Similarly, \tool marks the \emph{vault} and \emph{authority} accounts because their public keys are compared to the data of another account which also did not sign the transaction.
The oracle subsequently detects that the program writes a decreased amount of lamports in account \emph{vault}'s lamports field.
Since the account is marked as \emph{vulnerable} and is not previously compared to at least one public key of an account which signed the transaction, the oracle signals the VM.
However, after enabling the signer check in~\Cref{line:signercheck} of~\Cref{lst:msc}, \tool transitively marks all three accounts as non-vulnerable.

\begin{algorithm}
\caption{Missing Signer Check Oracle}
\label{algorithm:missing_signer_check}
\begin{algorithmic}[1]
        \State $k \gets \text{amount of accounts}$                          \label{missing_signer_check_1}
        \State $V \gets [\emptyset_1,\emptyset_2,...\emptyset_k]$           \label{missing_signer_check_2}
        \State $T \gets \emptyset$                                          \label{missing_signer_check_3}
        
        \While{program running}
            
            \If{program performs register comparison}                       \label{missing_signer_check_8}
                \State $a \gets \text{account in } src \text{ register}$    \label{missing_signer_check_9}
                \State $b \gets \text{account in } dst \text{ register}$    \label{missing_signer_check_10}
                \State $i \gets \text{index of account } a \text{ in } V$   \label{missing_signer_check_11}

                \If{$a$.data \textbf{is compared with} $b$.pubkey \textbf{and} $a \notin T$}    \label{missing_signer_check_12}
                    \If{\textbf{not} $a$ signed tx \textbf{and} \textbf{not} $b$ signed tx}                                      \label{missing_signer_check_13}
                        \State $V[i] \gets V[i] \cup \{a\} \cup \{b\}$                 \label{missing_signer_check_14}
                    \Else
                        \ForEach{$c \in V[i]$}                              \label{missing_signer_check_15}
                            \State $T \gets T \cup \{c\}$                   \label{missing_signer_check_16}
                        \EndFor
                        \State $V[i] \gets \emptyset$                       \label{missing_signer_check_17}        
                    \EndIf   
                \EndIf
                
            \EndIf
            
            \If{program writes to lamports or data of account $a$}          \label{missing_signer_check_18}
                \ForEach{$v \in V$}                                         \label{missing_signer_check_19}
                    \If{$a \in v$}                                          \label{missing_signer_check_20}
                        \State signal VM                                    \label{missing_signer_check_21}
                    \EndIf
                \EndFor
            \EndIf

        \EndWhile
\end{algorithmic}
\end{algorithm}

\paragraph{Missing Owner Check Oracle}
The \emph{missing owner check oracle} checks whether an eBPF \texttt{CMP} instruction compares data of an account that is \emph{not owned} by the program with the public key of another account.
In case the program performs such a comparison, the oracle marks the latter account as potentially vulnerable to a missing owner check, because the former account could be an attacker-controlled account in which the attacker could assemble data in such a way to reference arbitrary accounts.
\Cref{algorithm:missing_owner_check} details the bug-detecting process of the \emph{missing owner check oracle}.
The oracle operates as follows:
Before the program starts, the oracle initializes \emph{i)} a set $M$ in which it stores potentially malicious accounts (see Line \ref{missing_owner_check_1}) and \emph{ii)} a set of accounts $V$ that an attacker could potentially exploit using a missing owner check vulnerability (see Line \ref{missing_owner_check_2}).

At runtime of the program, the oracle checks if the program reads the data of an account $a$ which is not owned by the program, i.e., whose public key in the owner field is not equal to the program ID of the program (see Line \ref{missing_owner_check_3} and \ref{missing_owner_check_4}).
If the program does not own account $a$, the oracle marks $a$ as potentially malicious by adding it to set $M$ (see Line \ref{missing_owner_check_5}).

Moreover, the oracle checks at runtime whether the program compares account details of two account $a, b$ in registers (see Line \ref{missing_owner_check_6} to \ref{missing_owner_check_8}). 
In this comparison, if data from an account $a$ is compared to the public key of an account $b$, and $a$ is included in the set $M$ (see Line \ref{missing_owner_check_9}), the oracle marks $b$ as potentially vulnerable by adding $b$ to the set $V$ (see Line \ref{missing_owner_check_10}).
The same applies if the order of accounts in the register comparison is reversed, i.e., if the public key of account $a$ is compared with data of a account $b$ (see Line \ref{missing_owner_check_11} and \ref{missing_owner_check_12}). 
Thus, the oracle memorizes that the program has compared the data of an account potentially controlled by the attacker with the public key of another account.

The oracle signals the eBPF VM in case the program deducts lamports of an account which is contained in the set $V$, i.e., whose public key was previously compared with data of an account not owned by the program (see Line \ref{missing_owner_check_13} to \ref{missing_owner_check_17}).
The same applies when the program writes to the data field of an account contained in the set $V$ (see Line \ref{missing_owner_check_18} and \ref{missing_owner_check_19}).
This is because the oracle assumes that an owner check is missing because \emph{i)} the program compared the data of an account $a$ possibly controlled by the attacker with the public key of another account $b$ \emph{ii)} and then modified the lamports or data field of $b$.
An owner check would have already crashed the program, because the owner of the account whose data the program compared with a public key is not the program.

\begin{algorithm}
\caption{Missing Owner Check Oracle}
\label{algorithm:missing_owner_check}
\begin{algorithmic}[1]
        \State $M \gets \emptyset$           \label{missing_owner_check_1}
        \State $V \gets \emptyset$           \label{missing_owner_check_2}
        
        \While{program running}
            
            \If{program reads data of account $a$}                          \label{missing_owner_check_3}
                \If{$a$.owner $\neq$ program\_id}                           \label{missing_owner_check_4}
                    \State $M \gets M \cup \{a\}$                           \label{missing_owner_check_5}
                \EndIf
            \EndIf
            
            \If{program performs register comparison}                       \label{missing_owner_check_6}
                \State $a \gets \text{account in } src \text{ register}$    \label{missing_owner_check_7}
                \State $b \gets \text{account in } dst \text{ register}$    \label{missing_owner_check_8}

                \If{$a$.data \textbf{is compared with} $b$.pubkey \textbf{and} $a \in M$}       \label{missing_owner_check_9}
                    \State $V \gets V \cup \{b\}$                                               \label{missing_owner_check_10}
                \Else{ $a$.pubkey \textbf{is compared with} $b$.data \textbf{and} $b \in M$}    \label{missing_owner_check_11}
                    \State $V \gets V \cup \{a\}$                                               \label{missing_owner_check_12}
                \EndIf
            \EndIf
            
            \If{program writes to lamports field of account $a$}                \label{missing_owner_check_13}                   
                \State $l_{prev} \gets$ lamports of $a$ before write            \label{missing_owner_check_14}
                \State $l_{after} \gets$ lamports of $a$ after write            \label{missing_owner_check_15}
                \If{$a \in V$ \textbf{and}  $l_{after}$ $<$ $l_{prev}$}         \label{missing_owner_check_16}
                    \State signal VM                                            \label{missing_owner_check_17}
                \EndIf
            \EndIf
            
            \If{program writes to data of account $a$ \textbf{and} $a \in V$}   \label{missing_owner_check_18}                           
                \State signal VM                                                \label{missing_owner_check_19}
            \EndIf

        \EndWhile
%    \EndFunction
\end{algorithmic}
\end{algorithm}

Consider our example from~\Cref{lst:msc}.
Note that the code includes~\Cref{line:ownercheck} to show a valid owner check, i.e., in the example we assume that the code \emph{does not contain} the check at~\Cref{line:ownercheck}.
Due to the missing check, the oracle marks the \emph{vault} and \emph{authority} accounts as vulnerable, because in \Cref{line:check2,line:check3} their keys are compared to data read from the \emph{wallet\_info} account.
As \Cref{lst:msc} does not check the ownership of the \emph{wallet\_info} account, an attacker can forge this account such that his own public key assumes authority over the funds stored in any given \emph{vault} account.
To facilitate this, \tool uses the \emph{attacker-controlled accounts generator} (cf. \Cref{sec:blockchain-init}) to create attacker-controlled accounts that have references to accounts managed by the program as well as to the attacker to trigger operations on the accounts the program manages without verifying the owner of the account.
In this example, the attacker constructs the \emph{wallet\_info} account to contain his public key in the \emph{wallet.authority} field and the public key any given \emph{vault} account managed by the program in the \emph{wallet.vault} field.

Lastly, the oracle signals the VM as soon as an eBPF \texttt{STORE} instruction writes to the lamports or data field of any vulnerable account.
Note that when writing to a lamports field, the oracle signals the VM only if the \texttt{STORE} instruction decreases the account's lamports.
In our example, the oracle signals the VM at \Cref{line:vlamports}, because the program debits the \emph{vault} account, which is marked as vulnerable.

\paragraph{Lamports-based Oracle}
The \emph{lamports-based oracle} checks for user-related accounts that could lose lamports to an attacker-related account.
We illustrate the process for the \emph{lamports-based oracle} in~\Cref{algorithm:lamports_based_oracle}.
Before executing the program, the runtime initializes user-related and attacker-related accounts with PDAs (cf. \cref{sec:blockchain-init}) (see Line~\ref{lamports_based_oracle_1} and~\ref{lamports_based_oracle_2}).

This oracle is executed only if the issuer and signer of the transaction is the attacker (see Line~\ref{lamports_based_oracle_5}). 
Consider a program where the attacker passes user-related PDA seeds as a parameter:
An attacker could transfer lamports from user-related accounts to attacker-related accounts by specifying arbitrary seeds, despite proper owner and signer checks.

The lamports-based oracle checks eBPF \texttt{STORE} instructions to the lamports fields of accounts (see Line~\ref{lamports_based_oracle_6}).
Thus, the oracle checks each time if the program subtracts lamports from a user-related account (see Line~\ref{lamports_based_oracle_9} and \ref{lamports_based_oracle_10}) and credits lamports to an attacker-related account (see Line~\ref{lamports_based_oracle_11} and \ref{lamports_based_oracle_12}).
The oracle signals the VM if a program transfers lamports from a user-related account to an attacker-related account (see Line~\ref{lamports_based_oracle_13} and \ref{lamports_based_oracle_14}).
The lamports-based oracle only reports an error if neither the missing owner check, nor the missing signer check reports an error, since both oracles already check the lamports field.

\begin{algorithm}
\caption{Lamports-based Oracle}
\label{algorithm:lamports_based_oracle}
\begin{algorithmic}[1]
\State $U \gets \text{user related accounts}$                         \label{lamports_based_oracle_1}
\State $A \gets \text{attacker related accounts}$                     \label{lamports_based_oracle_2}
\State $\text{lamports}_{\text{user}}^{\text{lose}} \gets false$                        \label{lamports_based_oracle_3}
\State $\text{lamports}_{\text{attacker}}^{\text{gain}} \gets false$                  \label{lamports_based_oracle_4}
\While{program running \textbf{and} attacker signed tx}               \label{lamports_based_oracle_5}
    \State
    \If{program writes to lamports field of account $a$}                \label{lamports_based_oracle_6}                   
        \State $l_{prev} \gets$ lamports of $a$ before write \label{lamports_based_oracle_7}
        \State $l_{after} \gets$ lamports of $a$ after write        \label{lamports_based_oracle_8}

        \If{$a \in U$ \textbf{and}  $l_{after}$ $<$ $l_{prev}$}         \label{lamports_based_oracle_9}
            \State $\text{lamports}_{\text{user}}^{\text{lose}} \gets true$             \label{lamports_based_oracle_10}
        \ElsIf{$a \in A$ \textbf{and} $l_{after}$ $>$ $l_{prev}$}       \label{lamports_based_oracle_11}
            \State $\text{lamports}_{\text{attacker}}^{\text{gain}} \gets true$       \label{lamports_based_oracle_12}
        \EndIf
    \EndIf
    \State
    \If{$\text{lamports}_{\text{user}}^{\text{lose}} = \text{lamports}_{\text{attacker}}^{\text{gain}} = true$}                    \label{lamports_based_oracle_13}
        \State signal VM                                             \label{lamports_based_oracle_14} 
    \EndIf
\EndWhile
\end{algorithmic}
\end{algorithm}

\paragraph{Arbitrary CPI Oracle}
The \emph{arbitrary cross program invocation oracle} checks whether an eBPF \texttt{CALL} instruction calls the Solana-specific syscalls for invoking other programs using CPI.
In case the program executes such an instruction, the program checks whether a previously specified malicious key was passed to the syscall as the target program, which implies that an attacker is able to invoke his malicious program.
If a previously specified malicious key was passed as a target program to the syscall, the oracle signals the VM.

The oracle operates as follows:
Before starting the program, we define a malicious public key, which represents a malicious program controlled by the attacker (see Line \ref{arbitrary_cpi_line_1}).
At runtime of the program, the oracle then checks whether the program calls the functions \texttt{sol\_invoke\_signed\_rust()} or \texttt{sol\_invoke\_signed\_c()} (see Line \ref{arbitrary_cpi_line_2}). 
If this is the case, the oracle determines the public key of the called program (see Line \ref{arbitrary_cpi_line_3}). 
The oracle then checks if the public key of the program being invoked is equal to that of the previously defined malicious program (see Line \ref{arbitrary_cpi_line_4}).
This implies that an attacker is able to invoke his malicious program. 
If this is the case, the oracle signals the virtual machine (see Line \ref{arbitrary_cpi_line_5}).

\begin{algorithm}
\caption{Arbitrary CPI Oracle}
\label{algorithm:arbitrary_cpi}
\begin{algorithmic}[1]
\State $m_{pk} \gets \text{predefined malicious public key}$                                \label{arbitrary_cpi_line_1}
\While{program is running}
    \If{program calls \texttt{sol}\_\texttt{invoke}\_\texttt{signed}\_\texttt{rust()} \textbf{or} \texttt{sol}\_\texttt{invoke}\_\texttt{signed}\_\texttt{c()}}     \label{arbitrary_cpi_line_2}
        \State $i_{pk} \gets $ public key of invoked program                             \label{arbitrary_cpi_line_3}
        \If{$i_{pk} = m_{pk}$}                                                              \label{arbitrary_cpi_line_4}
                \State signal VM                                                            \label{arbitrary_cpi_line_5}
        \EndIf
    \EndIf
\EndWhile
\end{algorithmic}
\end{algorithm}

\paragraph{Missing Key Check Oracle}
The \emph{missing key check oracle} requires specifying the following two parameters before starting fuzzing: 
\emph{i)} the address of a function $f$ for which a key check is expected \emph{ii)} and an account \emph{a}, where the program is expected to compare the public key of an account it passes to function $f$ with the public key of account \emph{a} before calling function $f$.
In the case of Wormhole, $f$ would represent the function \emph{load\_instruction\_at} and \emph{a} would represent the \emph{KeccakSecp256k sysvar} account.
At program runtime, the oracle checks whether an eBPF \texttt{CALL} instruction calls the function $f$ and signals the VM if the program does not compare the public key of the account passed to the function $f$ with the public key of account \emph{a} before calling function $f$.

\Cref{algorithm:missing_key_check} details the bug-detecting process of the \emph{missing key check oracle}.
Before the program starts, we pass two parameters to the oracle: 
First, the address of a function $f$ for which a key check is expected (see Line \ref{missing_key_check_line_1}).
Second, the public key $a_{pk}$ of an account, where the program is expected to compare the public key of an account it passes to function $f$ with the public key $a_{pk}$ before calling function $f$ (see Line \ref{missing_key_check_line_2}).
In the case of Wormhole, $f$ would represent the function \emph{load\_instruction\_at} and $a_{pk}$ would represent the public key of the \emph{KeccakSecp256k sysvar} account.
Moreover, the oracle defines an empty list $A$, which it uses at program runtime to store accounts which public key was compared to $a_{pk}$ (see Line \ref{missing_key_check_line_3}). 

At program runtime, the oracle checks whether the program performs a register comparison operation or a function call operation.
In the case of the register comparison operation, the oracle uses the taint tracking engine to check whether the source and/or destination registers contain account details (see Line \ref{missing_key_check_line_5} and \ref{missing_key_check_line_6}). 
If at least one of the registers contains account details, the oracle checks whether the account details are the public key of an account and whether it is compared with $a_{pk}$ (see Line \ref{missing_key_check_line_7} and \ref{missing_key_check_line_9}). 
If this is the case, the oracle adds the account of the account details to the set $A$ to memorize that the account was compared with $a_{pk}$ before calling the function $f$ (see Line \ref{missing_key_check_line_8} and \ref{missing_key_check_line_10}).

In case the program calls the function $f$ using an eBPF \texttt{CALL} instruction (see Line \ref{missing_key_check_line_11}), the oracle extracts the account $p$ passed to the function (see Line \ref{missing_key_check_line_12}). 
Then the oracle checks if the account $p$ is not included in the set $A$ and thus was not compared with the public key $a_{pk}$ before calling the function $f$ (see Line \ref{missing_key_check_line_13}). 
If this is the case, the oracle signals the VM (see Line \ref{missing_key_check_line_14}).

\begin{algorithm}
\caption{Missing Key Check Oracle}
\label{algorithm:missing_key_check}
\begin{algorithmic}[1]
\State $f \gets \text{function address for which a key check is expected}$                                \label{missing_key_check_line_1}
\State $a_{pk} \gets \text{expected public key to be checked}$                     \label{missing_key_check_line_2}
\State $A \gets \emptyset$                                              \label{missing_key_check_line_3}

\While{program is running}

    \If{program performs register comparison}                           \label{missing_key_check_line_4}
        \State $a \gets \text{account in } src \text{ register}$   \label{missing_key_check_line_5}
        \State $b \gets \text{ account in } dst \text{ register}$   \label{missing_key_check_line_6}
        \If{$a_{pk}$ \textbf{is compared with} $b$.pubkey}        \label{missing_key_check_line_7}
            \State $A \gets A \cup \{b\}$                         \label{missing_key_check_line_8}
        \Else{ $a$.pubkey \textbf{is compared with} $a_{pk}$}     \label{missing_key_check_line_9}
            \State $A \gets A \cup \{a\}$                         \label{missing_key_check_line_10}
        \EndIf
    \EndIf

    \If{program calls function $f$}                                     \label{missing_key_check_line_11}
        \State $p \gets \text{passed account to function } f$                   \label{missing_key_check_line_12}
        \If{$p \notin A$}                                               \label{missing_key_check_line_13}
            \State signal VM                                            \label{missing_key_check_line_14}
        \EndIf
    \EndIf

\EndWhile
\end{algorithmic}
\end{algorithm}

\paragraph{Integer Bugs Oracle}
The \emph{integer bugs oracle} checks whether an eBPF \texttt{STORE} instruction writes an overflowed or underflowed value to the lamports field of an account. 
Subsequently, the oracle signals the VM.
To do so, this oracle leverages the taint tracking engine.
Note that the oracle is not applicable to the accounts' data field, since the Borsh serializer\footnote{\url{https://borsh.io/}}---which programs frequently use for serializing and deserializing account data---calculates overflow or underflow values when serializing data into the account's data field. 

\Cref{algorithm:integer_bugs} provides a detailed explanation of the \emph{integer bug oracle}.
The oracle operates as follows:
Before the program runs, the oracle initializes an empty list $T$ in which the oracle stores the registers that contain bugged values, i.e., overflowing or underflowing values (see Line \ref{integer_bugs_line_1}). 

At program runtime, the oracle then checks whether an eBPF \texttt{STORE} instruction has caused the program to write a value to a register (see Line \ref{integer_bugs_line_2}). 
If this case occurs, the oracle checks whether the value which the program writes into the register is bugged, i.e. has overflowed or underflowed (see Line \ref{integer_bugs_line_3}).
If a bugged value is written to the register, the oracle remembers the register by adding the register to the list $T$ (see Line \ref{integer_bugs_line_3} and \ref{integer_bugs_line_4}).
In case the program does not write a bugged value to the register, the oracle removes the register from the list $T$ (see Line \ref{integer_bugs_line_5}).

Moreover, the oracle checks if the program writes a value into the lamports field of an account at runtime (see Line \ref{integer_bugs_line_6}).
Then the oracle checks whether the set $T$ contains the register whose value the program writes to the lamports field of an account (see Line \ref{integer_bugs_line_7} and \ref{integer_bugs_line_8}).
This indicates that the program has written a bugged value to the lamports field of the account.
If the program writes a bugged value into the field, the oracle signals the VM (see Line \ref{integer_bugs_line_9}).

\begin{algorithm}
\caption{Integer Bugs Oracle}
\label{algorithm:integer_bugs}
\begin{algorithmic}[1]
\State $T \gets \emptyset$                                  \label{integer_bugs_line_1}
\While{program is running}
    \If{program writes $value$ to $src$ register}           \label{integer_bugs_line_2}
        \If{$value$ is bugged}                              \label{integer_bugs_line_3}
            \State $T \gets T \cup \{src\}$                 \label{integer_bugs_line_4}
        \Else
            \State $T \gets T \backslash \{src\}$           \label{integer_bugs_line_5}
        \EndIf    
    \EndIf
    
    \If{program writes $value$ to lamports field}           \label{integer_bugs_line_6}
        \State $src \gets \text{register holding } value$   \label{integer_bugs_line_7}
        \If{$src \in T$}                                    \label{integer_bugs_line_8}
            \State signal VM                              \label{integer_bugs_line_9}
        \EndIf
    \EndIf
\EndWhile
\end{algorithmic}
\end{algorithm}

\subsubsection{Program Semantic Extractors}
\label{sec:program-semantic-extractors}

\tool does not have access to a program's source code, and therefore RunDelSol implements a \emph{PDA seed structure extractor} and an \emph{account data structure extractor} to determine semantic program information.

\paragraph{PDA Seed Structure Extractor}
The PDA seed structure extractor extracts the structure and origin of the seeds used by the program to derive PDAs.
For this purpose, we use the taint tracking engine to determine the origin of the arguments of Solana-specific syscalls that derive PDAs.
This allows the PDA seed structure extractor to determine, for example, whether a seed originates from an account's public key.
In case no origin for a seed is available, we assume that the seed has been statically compiled into the program. 
Hence, the PDA seed structure extractor does not record the origin of the seed (e.g., a public key) but the actual bytes of the seed.

\paragraph{Account Data Structure Extractor}
The account data structure extractor extracts positions of public keys in the data structures of accounts.
The extractor checks whether the program writes a public key to the data field of an account located on the \emph{program input} (cf. \Cref{fig:program_input_layout}).
In \Cref{lst:msc}, the \emph{account data structure extractor} extracts the offsets of the \emph{wallet.authority} and \emph{wallet.vault} fields in the \emph{wallet\_info} account.
Note that this information is extracted during account initialization of the \emph{wallet\_info} account. 
Further, \tool leverages this information in subsequent fuzzing runs to populate the data fields of \emph{wallet\_info}-like accounts with existing public keys that may refer to either attacker-controlled accounts or program-managed accounts.
Regarding the program in \Cref{lst:msc}, this means that \tool creates a \emph{wallet\_info} account with the exact keys required to trigger a bug that leads to illegitimate gain of lamports.
Thus, the information persists across the independent transactions created by the fuzzer.
The \blockemu uses this information to deploy accounts on the blockchain which can trigger potential missing owner check vulnerabilities. 

\subsection{Transaction Evaluator}\label{sec:transaction-eval}
After RunDelSol finishes the execution of the transaction, RunDelSol forwards the coverage information to the transaction evaluator.
This includes the PDA seed structures, account data structures, and---if a bug was detected---the oracle signals, received at runtime.
The coverage information and oracle signals determine the input bytes of the next fuzzing iteration.
If the oracle signaled the eBPF VM, the transaction evaluator also creates a vulnerability report, that includes information to reproduce the transaction. 
In case RunDelSol extracted new PDA seed structures or account data structures, the transaction evaluator informs the \blockemu to generate a new \ledgersnap before starting the next fuzzing iteration.
\section{Evaluation}

In this section, we evaluate multiple aspects of \tool on several datasets of Solana programs.
We start by demonstrating the soundness and completeness precision of \tool with the \emph{Neodyme Breakpoint Workshop} dataset~\cite{Neodyme2021-vw} in~\Cref{sec:eval_valid}.
This dataset is a collection of prevalent Solana program vulnerabilities and is used in previous work~\cite{Cui2022-nm}.
Furthermore, we compare \tool with VRust~\cite{Cui2022-nm}, which is currently the only other approach addressing Solana program security.
Second, we test \tool's vulnerability discovery effectiveness on real-world programs directly taken from the Solana mainnet blockchain. We present our findings and discuss newly discovered bugs in~\Cref{sec:eval_bugs}.
Lastly, in~\Cref{sec:eval_perf}, we demonstrate \tool's performance with bug bounty programs~\cite{immunefi}.
We focus on \tool's test case throughput and achieved code coverage.

\paragraph{Experimental Setup} 
We ran our evaluation on an AMD EPYC 7302P CPU with 16 cores clocked at \SI{3}{GHz} with \SI{256}{GB} RAM.
The experiments are executed in parallel, keeping all physical CPU cores fully occupied.
Each fuzzing experiment uses a single core and uses the same initial seed for all fuzzing runs.

\subsection{Bug Detection Capabilities}
\label{sec:eval_valid}

To validate our design, we test \tool with the Neodyme Breakpoint Workshop dataset~\cite{Neodyme2021-vw}.
The dataset contains common Solana vulnerabilities (cf.~\cref{sub:sol-vulnerabilities}).
This dataset is organized into 5 different levels, where each level consists of a Solana program with a specific vulnerability. 
For this experiment, we fuzz each program with a timeout of 10 minutes.
Our results are depicted in~\Cref{tbl:validity}:
\tool is able to find all the bugs in this dataset within less than 5 seconds.
\tool does \emph{not report any false alarms}, and is able to precisely detect each vulnerability.

We do not include the \emph{Level}~\emph{3} program, because the program has an \emph{account confusion vulnerability}.
Detecting account confusions requires knowledge of the underlying data layout that represents the expected data structure in memory.
This information requires access to the source code of the program.
However, \tool's goal is to detect bugs in Solana programs, without relying on source code, and thus we skip this program.

\paragraph{Comparison with VRust}
In contrast to VRust~\cite{Cui2022-nm}, \tool reliably detects bugs in smart contracts, without source code.
Thus, a full comparison of every metric (e.g., performance) with VRust is impossible.
While VRust is able to detect the same bugs as \tool, it reports false alarms regarding the integer bug in the \emph{Level}~\emph{2} program.
Meanwhile, \tool does not report a single false alarm in this dataset.
In addition, VRust only indicates a missing key check in the \emph{Level}~\emph{0} program and \emph{Level}~\emph{1} program.
\tool, on the other hand, can trace the vulnerability to a missing owner check in the \emph{Level}~\emph{0} program and a missing signer check in the \emph{Level}~\emph{1} program, resulting in \tool being more precise compared to VRust.

%!TEX root = ../main.tex

% \begin{table}[]
% \centering
% \begin{tabular}{@{}l|ccccc|c@{}}
% \toprule
% \multirow{2}{*}{Program} & \multicolumn{5}{c|}{Vulnerabilities} & Time to first \\ 
%                   & MOC     & MSC   & IB    & ACPI & MKC &  Bug (s)      \\
% \midrule
% \texttt{Level 0}  & \tyes   &       & \tyes &     &  & \num{4}                   \\
% \texttt{Level 1}  &         & \tyes & \tyes &     &  & \num{2}                     \\
% \texttt{Level 2}  &         &       & \tyes &     &  & \num{2}                     \\
% %\texttt{Level 3}  &         &       &       &    &   &                       \\
% \texttt{Level 4}  &         &       &       & \tyes & &  \num{1}                     \\
% \midrule
% \texttt{Wormhole}$^{\ast}$  &         &       &       & & \tyes &  \num{37}                     \\
% \midrule

% \end{tabular}%
% \caption{Results of our validity measurement. We mark true bugs with \tyes.}
% \label{tbl:validity}
% \end{table}

\begin{table}[t]
\centering
\resizebox{\columnwidth}{!}{%
\begin{tabular}{@{}l|ccccc|c@{}}
\toprule 
\multirow{2}{*}{Program}    & \multicolumn{5}{c|}{Vulnerabilities}                                                                                     & Time to first        \\
                            & IB                         & MOC   & MSC                      & ACPI                      & MKC                         & Bug (s)              \\
\midrule 
\texttt{Level 0}            & \tyes                      & \tyes &                          &                           &                             & \num{4}              \\
\texttt{Level 1}            & \tyes                      &       & \tyes                    &                           &                             & \num{2}              \\
\texttt{Level 2}            & \tyes                      &       &                          &                           &                             & \num{2}              \\
\texttt{Level 4}            &                            &       &                          & \tyes                     &                             & \num{1}              \\
\midrule 
\texttt{Wormhole}$^{\ast}$  &                            &       &                          &                           & \tyes                       & \num{37}             \\
\midrule 
False Alarms (\tool{} : VRust) & \multicolumn{1}{c}{0 : 62} & 0 : 0 & \multicolumn{1}{c}{0 : 0} & \multicolumn{1}{c}{0 : 3} & \multicolumn{1}{c|}{0 : 107} & \multicolumn{1}{c}{N/A}
\end{tabular}%
}
\caption{Results of our validity measurement. We mark true bugs with \tyes.}
\label{tbl:validity}
% \vspace{-9ex}

\end{table}

\paragraph{The Infamous Wormhole Bug}
In February 2022, wrapped Ether (wETH) with a value of 323 million USD has been stolen from the Wormhole program, which implements a bridge between Ethereum and Solana.
The underlying bug is a missing key check in a Solana program.
\tool implements a bug detection oracle that detects this bug.
Furthermore, \tool works on a binary-only level and does not need to understand a program's semantics to detect vulnerabilities.
However, by design, the original Wormhole program and the underlying bug requires this level of context information.
The context is provided by off-chain guardians that check and verify each transaction.
However, we challenged \tool to detect this bug \emph{without} any context information.
Therefore, we created an emulation of this program which shares the same vulnerability as the Wormhole bug.
Here, \tool was able to detect the bug in less than 40 seconds.

\subsection{Discovering New Bugs}
\label{sec:eval_bugs}
To evaluate \tool's effectiveness in discovering unknown vulnerabilities, we assembled a dataset of \noc real-world programs deployed on the Solana mainnet on March 27, 2023.
We take the following steps to ensure that we have the most current and complete collection of Solana programs:
First, we query an RPC node of the Solana network for all programs that belong to the most recent loader program\footnote{At the time of writing, this is \emph{BPFLoaderUpgradeab1e11111111111111111111111}.}.
Second, we use the Solana toolchain to dump each of the programs into an ELF file.
VRust~\cite{Cui2022-nm} is not able to analyze any of these programs, which emphasizes the gap that \tool fills. 
Given the large data set of contracts, we set the timeout to 5 minutes.
In total, \tool reports \num{92} potential security vulnerabilities in \num{52} out of the \noc programs, including \num{30} \emph{missing signer checks}, \num{12} \emph{arbitrary CPIs}, and \num{30} \emph{integer bugs}.
Moreover, \num{20} reports indicate potential vulnerabilities to lamports theft without possessing the vulnerabilities listed before.

\begin{table}[t]
  \centering
  \begin{tabular}{@{}c|c@{}}
    \toprule
    Abbreviation & Full Program ID \\
    \midrule
    \texttt{3nJ2...5erP} & \texttt{3nJ2MWbnS3bW8rnWhejAgnLQTyiqoA5fMq5Z7jRv5erP} \\
    \texttt{3od3...jnsW} & \texttt{3od3X7QN84FTonkyXbQiT1ydxcT9P1jBcA9mbgD3jnsW} \\
    \texttt{3Vtj...4q8v} & \texttt{3VtjHnDuDD1QreJiYNziDsdkeALMT6b2F9j3AXdL4q8v} \\
    \texttt{3w57...obPW} & \texttt{3w57iMhv5Zk5VDuTe5dspm2FzE9zQhscCb9CZpAKobPW} \\
    \texttt{4hPk...JP5N} & \texttt{4hPkNV2WsgPW1wHHcQebvV7GLyLgdDDLEx3Pu6LzJP5N} \\
    \texttt{4M2f...jStx} & \texttt{4M2fancicHbUtMLcMNmbi97YngFoqBcnFk5D31JjjStx} \\
    \texttt{6Lan...szqi} & \texttt{6LanqAFCbucXWSG35ssij4kFDTWJ25BY7d6hbR2szqi} \\
    \texttt{7FWE...9p7p} & \texttt{7FWEcVG1YRW7evGR3bXgu47ge8m6Je7BQuvTzMbn9p7p} \\
    \texttt{9a5d...jZP9} & \texttt{9a5dihgNgBhWnjmRDJ8rUy4ihetvgMmjaPk7NGdsjZP9} \\
    \texttt{9tSW...11yy} & \texttt{9tSWsKwtDL6YseLuh1haGFJk312uu9HGyrnVa5XH11yy} \\
    \texttt{GQ6q...1u6K} & \texttt{GQ6qchUsofiK7rzeFg5jbvpHcJ7pNnfL4yfwaYrB1u6K} \\
    \texttt{9WoL...849B} & \texttt{9WoLnfjLKk1EBtkABhe3vcA8CLogsbs3XBoddn8h849B} \\
    \texttt{H5rp...nPSG} & \texttt{H5rpfCD6hLFCPCfxxqjGg94Gqoigqfk7afhqGLu1nPSG} \\
    \bottomrule
  \end{tabular}
  \caption{Program IDs and abbreviations from~\Cref{tbl:eval_bugs}.}%
  \label{tbl:full-addresses}
\end{table}

Confirming vulnerabilities is challenging due to the absence of source code. Hence, we opted for the following approach. We first generate instructions based on the payload information contained in the vulnerability report generated by \tool.
Next, we analyze the program logs as well as the disassembled eBPF bytecode executed at the runtime of the instructions.
Afterward, we craft transactions and observe if the transactions create an erroneous state in the blockchain. Note that this is a tedious validation process, but a common issue when developing smart contract fuzzers~\cite{echidna,sfuzz,rodler2023efcf}.

At the time of writing, we are able to validate the existence of 14 exploitable bugs, and 2 non-exploitable bugs.
Accordingly, \tool currently has a false alarm rate of \SI{12.5}\%.
\cref{tbl:eval_bugs} shows the 16 discovered bugs and public key abbreviations of the vulnerable programs.
For the sake of reproducibility, we list the full public keys in~\Cref{tbl:eval_bugs}.
In the following, we present five interesting vulnerabilities detected by our approach.
To minimize potential damage, we ensured that none of these programs are actively managing valuable assets and that no token accounts exist to which the programs are assigned as authority.

\paragraph{Responsible Disclosure and Ethical Concerns}
Since we conduct this experiment on all Solana programs present on the Solana blockchain, this includes many programs of unknown origin, i.e., the authors are anonymous.
We tried our best to reach out to the program authors of this experiment.
Due to the lack of contact information, we could not disclose our findings directly to the authors of vulnerable Solana programs.
Hence, we decided to disclose \emph{all} of our findings to the Solana foundation\footnote{\url{https://solana.org/}} and offered collaboration to fix the vulnerabilities.

\begin{table}[t]
\centering
\begin{tabular}{@{}l|cccc@{}}
\toprule
\multirow{2}{*}{Program ID} & \multicolumn{4}{c}{Vulnerabilities} \\
                         & MSC   & Integer Bug & ACPI  & Lamport \\
\midrule
% 3nJ2MWbnS3bW8rnWhejAgnLQTyiqoA5fMq5Z7jRv5erP: MSC FP
\texttt{3nJ2...5erP}     & \tno  &       &       &       \\
% 3od3X7QN84FTonkyXbQiT1ydxcT9P1jBcA9mbgD3jnsW: ACPI
\texttt{3od3...jnsW}     &       &       & \tyes &       \\
% 3VtjHnDuDD1QreJiYNziDsdkeALMT6b2F9j3AXdL4q8v: IB
\texttt{3Vtj...4q8v}     &       & \tyes &       &       \\
% 3w57iMhv5Zk5VDuTe5dspm2FzE9zQhscCb9CZpAKobPW: ACPI
\texttt{3w57...obPW}     &       &       & \tyes &       \\
% 4hPkNV2WsgPW1wHHcQebvV7GLyLgdDDLEx3Pu6LzJP5N: ACPI
\texttt{4hPk...JP5N}     &       &       & \tyes &       \\
% 4M2fancicHbUtMLcMNmbi97YngFoqBcnFk5D31JjjStx: IB, MSC
\texttt{4M2f...jStx}     & \tyes & \tyes &       &       \\
% 6LanqAFCbucXWSG35ssij4kFDTWJ25BY7d6hbR2szqi: MSC
\texttt{6Lan...szqi}     & \tyes &       &       &       \\
% 7FWEcVG1YRW7evGR3bXgu47ge8m6Je7BQuvTzMbn9p7p: ACPI
\texttt{7FWE...9p7p}     &       &       & \tyes &       \\
% 9a5dihgNgBhWnjmRDJ8rUy4ihetvgMmjaPk7NGdsjZP9: IB
\texttt{9a5d...jZP9}     &       & \tno &       & \tyes \\
% 9tSWsKwtDL6YseLuh1haGFJk312uu9HGyrnVa5XH11yy: IB, MSC
\texttt{9tSW...11yy}     & \tyes & \tyes &       &       \\
% GQ6qchUsofiK7rzeFg5jbvpHcJ7pNnfL4yfwaYrB1u6K: IB
\texttt{GQ6q...1u6K}     &       & \tyes &       &       \\
% 9WoLnfjLKk1EBtkABhe3vcA8CLogsbs3XBoddn8h849B
\texttt{9WoL...849B}     &  &       &   \tyes    &       \\
% H5rpfCD6hLFCPCfxxqjGg94Gqoigqfk7afhqGLu1nPSG: IB
\texttt{H5rp...nPSG}     &       & \tyes &       &       \\
% X3NongcyQZDhmvFYWbUASHXurHTTyPXYfdDEci99uuR: MSC (2)
%\texttt{X3No...9uuR}     & \tyes$^{\ast}$ &       &       &       \\
\midrule
\end{tabular}%
\caption{Results of our bug discovery experiment. We mark true bugs with \tyes and false alarms with \tno.}%\\$^{\ast}$} %Two distinct bugs of this type were investigated and are actual bugs.}
\label{tbl:eval_bugs}
\vspace{-2 em}
\end{table}

\finding{Integer bugs in 3Vtj...4q8v} This program contains an \emph{integer bug} that an attacker can use to steal lamports from a program-controlled account. 
The vulnerable instruction requires four accounts, and subtracts one SOL from the fourth account while adding it to the first account.
However, when reducing and crediting, the program does not check whether an overflow or underflow of lamports has occurred. 
Given that the first account owns too much SOL and the fourth account owns too little, an attacker could exploit the integer bug to add lamports to the fourth account and subtract lamports from the first account.

\finding{Arbitrary CPI in 3w57...obPW}
\tool found an \emph{arbitrary CPI} vulnerability.
We confirmed the bug by sending a malicious instruction.
The instruction accepts five accounts, where the last account signs the transaction.
The program then invokes the first account supplied without checking the account.

\finding{Arbitrary CPI in 9WoL...849B}
\tool detected an \emph{arbitrary CPI} vulnerability which grants additional privileges to the invoked program by signing a PDA in the CPI instruction.
The program \texttt{9WoL...849B} manages accounts whose public key follows a PDA seed structure consisting of one seed corresponding to a passable public key (e.g., a wallet account). 
Similarly to \texttt{3w57...obPW}, we have also created an instruction for \texttt{9WoL...849B} to confirm the bug:
The instruction expects \emph{13} accounts, where account \emph{13} is the program an attacker can invoke arbitrarily, and accounts \emph{3}, \emph{8}, and \emph{11} are accounts that have the same public key matching the PDA seed structure of \texttt{9WoL...849B}, i.e., a PDA associated to \texttt{9WoL...849B}.
The instruction results in \texttt{9WoL...849B} invoking the arbitrary invocable program and signing the PDA in the CPI instruction.
This leads to the invoked program having additional privileges than originally included in the transaction. 
Since the program states in the program log that it generates an instruction for calling the function \emph{borrow\_obligation\_liquidity} of the Serum Swap program before performing CPI, we assume that it is supposed to manage accounts of the Serum Swap program as an authority.

\finding{Multiple Vulnerabilities in 4M2f...jStx} 
Here, \tool discovered both a \emph{missing signer check} and an \emph{integer bug} in this program. 
The program expects an instruction containing four accounts and allows playing a gambling game in which a user can win or lose. 
The number of lamports of the second account multiplied by an odd of \emph{1.9934} determines the total payout of the game. 
In case the user wins the game, the program credits the payout to the first account and subtracts
\begin{inparaenum}[1)]
    \item from the fourth account the lamports worth \emph{0,9334} multiplied by the lamports of the second account, and 
    \item from the second account, its total lamport balance.
\end{inparaenum}
When crediting and subtracting the lamports, the program does not check whether overflows or underflows have occurred.
Hence, an attacker can exploit the integer bug to credit the fourth account with lamports instead of subtracting lamports in the case that the user won the game. 
In addition, the program does not check which account signed the transaction. 
Thus, an attacker can submit arbitrary program-controlled accounts and start gambling without the program ever checking whether the attacker is authorized to gamble with the submitted accounts. 
We note that the programs \texttt{6Lan...szqi} and \texttt{9tSW...11yy} also allow gambling similar to \texttt{4M2f...jStx}, and also do not verify that gambling with the submitted accounts is authorized. 

\finding{Integer Bug in 9a5d...jZP9}
Besides suffering from an integer bug, this program also enables an attacker to transfer lamports from a program-controlled account to an arbitrary, attacker-controlled account. 
The program expects two accounts and subtracts all lamports of the first account and credits them to the second account without checking for overflow or underflow of lamports. 
Thus, the integer bug is not exploitable, as only the lamports field of the second account can overflow but not of the first account. 
However, the program allows passing an arbitrary program-controlled account as a first account, transferring its lamports to the second account. 
In general, such a behavior is undesirable, as it allows an attacker to drain funds of all accounts belonging to this program.
This is a clear indication of an access control bug:
the intended program behavior would surely only allow an \emph{authorized} account to transfer lamports from program-controlled accounts to carefully selected accounts.  
To address this bug, the program must verify that the authorized account is included in the instruction and that it signed the transaction to prevent exploitation. 

\subsection{Performance Analysis}
\label{sec:eval_perf}

Given that \tool is the first fuzzer for Solana programs, there exists no qualitative baseline or dataset to measure common fuzzing metrics like coverage and execution speed.
Thus, we also aim at establishing a baseline allowing the community to compare future fuzzers with \tool.

We assembled a dataset from the Immunefi bug bounty list~\cite{immunefi}. 
This dataset includes a diverse set of Solana programs used in production and offers a higher grade of code quality and complexity, compared to the average mainnet programs. 
Hence, we assess the execution speed and code coverage of \tool based on this dataset.
For this experiment, we use a representative timeout of 24 hours and collect metrics on execution speed and code coverage.

\paragraph{Binary-only Approach Baseline}
Unlike the experiments in \Cref{sec:eval_bugs}, source code for the bug-bounty dataset is available.
The performance and coverage of \tool could potentially be optimized by analyzing the source code to extract context information about authority or configuration accounts, solving assertions, and uncovering new code paths. 
However, we refrain from doing so because (1)~we aim to provide representative measurements and (2) for the large majority of Solana programs no source code is available. 

\paragraph{Challenge: Measuring Code Complexity of Solana programs}
There is currently no tool support to measure the complexity of Solana smart contracts.
Previous work, like VRust~\cite{Cui2022-nm}, relied on lines of code (LOC) to estimate the complexity of a program. However, this metric is insufficient since it includes unreachable code.
Furthermore, \tool's coverage works on traversed edges in a program's control flow graph (CFG) and is therefore incomparable to LOC.
To tackle this challenge, we developed a static analysis approach which measures the complexity based on traversing the eBPF code and counting every control flow instruction that eBPF supports.
We use this number to over-approximate the number of edges in the CFG.

Note that this ensures that we include every eBPF \texttt{JMP},
\texttt{CALL}, and \texttt{RET} instruction.
This includes any edges that are by design not reachable, since these may represent dead code or Solana-specific error handling routines.
For example, one routine is the handling of an incorrect serialized program id at the program input, which is never executed due to the valid instructions generated by the transaction generator.

We argue that this is sufficient to estimate the complexity of Solana programs because it provides a better insight into the complexity of a program than LOC.
We analyze the target contracts with our CFG-based approach, and compare the results with the covered code paths by \tool.

%!TEX root = ../main.tex

\begin{table}[t]
\centering
\begin{adjustbox}{max width=\linewidth}
\begin{tabular}{@{}l|r|rr|r@{}}%|rrr@{}}
\toprule
\multicolumn{1}{c|}{Program} & \multicolumn{1}{c|}{Bounty}   & \multicolumn{1}{r}{\#CFG} & \multicolumn{1}{r|}{Covered} & \multicolumn{1}{r}{Mean} \\ %& \multirow{2}{*}{Alarms} & \multicolumn{1}{c}{Verified} & Time to first \\
\multicolumn{1}{c|}{Name}    & \multicolumn{1}{c|}{(\$)}     & \multicolumn{1}{r}{Edges}  & \multicolumn{1}{r|}{Edges} &  \multicolumn{1}{r}{Tx/s} \\%     &           & \multicolumn{1}{c}{Alarms} & Bug (s) \\
\midrule
\texttt{Drift Protocol}           & \num{500000}  & \num{67552} & \num{2336} & \num{5211} \\%& 0 & --- & ---\\
\texttt{Jet Airspace}             & \num{100000}  & \num{7559}  & \num{1398} & \num{1951} \\%& 0 & --- & ---\\
\texttt{Jet Control}              & \num{100000}  & \num{9506}  & \num{2716} & \num{1500} \\%& 0 & --- & ---\\
\tno \texttt{Jet Fixed Term}   & \num{100000}  & \num{27246} & \num{2375} & \num{1256}   \\%& 1 & 0   & 115\\         % FP
\texttt{Jet Margin}               & \num{100000}  & \num{18216} & \num{1332} & \num{1771} \\%& 0 & --- & ---\\
\texttt{Jet Margin Swap}          & \num{100000}  & \num{12843} & \num{2626} & \num{1332} \\%& 0 & --- & ---\\
\texttt{Jet Metadata}             & \num{100000}  & \num{5164}  & \num{1310} & \num{786}  \\% & 0 & --- & ---\\
\tyes \texttt{Jet Test Service}   & \num{100000}  & \num{17972} & \num{2706} & \num{2894} \\%& 1 & 1   & 1  \\
\texttt{Lido}                     & \num{2000000} & \num{8731}  & \num{1305} & \num{274}  \\%& 0 & --- & ---\\           % Anker: same
\texttt{Marinade Finance}         & \num{250000}  & \num{22424} & \num{2648} & \num{1023} \\%& 0 & --- & ---\\
\texttt{Port Finance VRL}         & \num{500000}  & \num{10704} & \num{1643} & \num{1066} \\%& 0 & --- & ---\\
\texttt{Pyth}                     & \num{500000}  & \num{4438}  & \num{2984} & \num{576}  \\%& 0 & --- & ---\\
\texttt{Solend Program}           & \num{1000000} & \num{10818} & \num{1681} & \num{1129} \\%& 0 & --- & ---\\
\texttt{Sundial}                  & \num{500000}  & \num{16792} & \num{3171} & \num{1896} \\%& 0 & --- & ---\\
\texttt{Token Faucet}             & \num{500000}  & \num{6500}  & \num{1299} & \num{1217} \\%& 0 & --- & ---\\
\texttt{Whirlpool}                & \num{500000}  & \num{20593} & \num{2726} & \num{1229} \\%& 0 & --- & ---\\
\midrule
\emph{$n=16$}                    &                & $\overline{x} ={16691}$  & $\overline{x}={2141}$ & $\overline{x}={1569}$\\
%\emph{$n=16$}                    &                & $\sum ={16691}$  & $\overline{x}={ }???$ & $\overline{x}={ }???$\\
\bottomrule
\end{tabular}%
\end{adjustbox}
\caption{Results of our performance experiment.}% We mark programs where \tool reports a valid bug with \tyes. We use the \tno marker for reported bugs that we cannot validate.}
\label{tbl:performance}
\vspace{-2 em}
\end{table}

\paragraph{Coverage}
\Cref{tbl:performance} shows the results of this experiment.
First, we observe that the estimated complexity of the dataset varies widely, ranging from \num{4438} edges in the control flow graph to \num{67552}.
This confirms that the size and complexity of the programs in the dataset is diverse.
The number of covered edges by \tool ranges from \num{1299} to \num{3171}.
This provides two important insights: First, \tool is able to consistently generate meaningful transactions to uncover new program paths.
Second, the binary-only analysis approach leads to a number of programs, where the number of covered edges does not increase over time.
By further investigation, we learned that certain barriers or assertions prevent \tool from reaching deeper nested code.
Hence, there is room for optimization for future work. For example, extending \tool with symbolic execution~\cite{fuzz-symex, Mossberg2019-xp}, using Redqueen~\cite{aflpp, Aschermann2019-ha}, or, as mentioned before, incorporating the available source code (\Cref{sec:eval_bugs}), to overcome these roadblocks.

\paragraph{On Fuzzing Throughput}
Another interesting insight is that \tool is capable to generate on average more than 1569 transactions per second for every Solana program.
Furthermore, \tool has an average of over \num{1000} transactions per second for 13 out of 16 programs. 
However, even in these three outliers, \tool is able to generate at minimum \num{274} transactions per second.
We understand that \tool regularly extracts new runtime semantics for these programs, causing the blockchain emulator to initiate updating the \ledgersnap as well as deriving new PDAs.
We measured that initializing the snapshot takes about \num{60}ms, which results in fewer transactions per second being generated from the input bytes.

\paragraph{Vulnerability Reports}
\tool reported 2 bugs in the bug dataset both belonging to the Jet Protocol. 
We investigate the bugs while also consulting the source code of each program. As we will see, one of the bugs is a false positive while we believe the other one is a true positive which is currently under review by the developers.

\finding{False Alarm in Jet Fixed Term}
\tool reports a missing signer check in the \texttt{Jet Fixed Term} program, which is a program for fixed-term lending and borrowing.
The missing signer check exists in a function that cancels orders, which requires two accounts as its input.
While the first account strictly belongs to a user of this program, the second account is a public order book containing the orders.
By signing the transaction, a user is granted authority to remove an order from the marketplace.
As a reminder, the oracle of \tool (c.f.~\Cref{sub:oracle}) considers transitive signer checks if accounts are linked in some way, i.e., an account $a$ may refer to another account $b$ if $a$'s data field contains the public key of $b$.
Additionally, the public key stored in $a$ must be compared to the public key of $b$.
However, in this particular case, the order book's account data may contain the public key of the authority account, but the public key of the authority account---which signed the transaction---is never compared to it.
As a result, \tool reports a missing signer check for this program.
However, we consider this a false alarm, as the bug is not exploitable, because the program checks whether the user account owns the canceled order.

\finding{True Positive Bug in Jet Test Service}
\tool reports an arbitrary cross-program invocation for the \texttt{Jet Test Service} program from the Jet Protocol.
The arbitrary cross-program invocation exists in a function that accepts an arbitrary amount of accounts as long as a minimum of two accounts is provided:
the first account is a potentially uninitialized account, and the second account can be any other program of the Solana blockchain.

The function then checks whether the first account provided is initialized on the Solana blockchain, and if this is \emph{not} the case, the function invokes the second program using CPI.
Since there are no restrictions on the choice of account to invoke, we consider this a true bug in the \texttt{Jet Test Service} program.
This bug can be overcome by having the \texttt{Jet Test Service} program verify the public key of the second account before invoking the second account using CPI. 

We are now in contact with the vendor to fix these issues and confirm the bugs.
In conclusion, we can confirm the ability to fuzz complex targets with high transaction throughput and coverage.
%!TEX root = ../main.tex

\section{Related Work}

In this section, we survey additional recent research work in the area of detecting bugs in smart contracts and fuzzing. 

\paragraph{Solana Bug Detection Approaches}
To the best of our knowledge, VRust~\cite{Cui2022-nm} is currently the only static analysis tool for Solana programs.
VRust finds missing owner, signer, and key checks, integer, account confusion, cross program invocation, numerical precision error, and bump seed bugs.
It detects these bugs by analyzing source code and checking for vulnerable patterns.
However, VRust relies on the availability of source code which is unfortunately not available for the majority of Solana programs.
Moreover, in contrast to \tool, VRust suffers from a high false alarm rate of \num{89,58}\%.
Hence, due to the overwhelming number of alarms, it is very likely that developers will miss the true positives.

\paragraph{Library Fuzzing}
Fuzzing~\cite{Bohme2016-xt, Bohme2017-vv, Liang2018-ws, nyx, aflpp, libafl} is a popular technique for evaluating the security of software and hardware components~\cite{Feng2020-ni, Maier2020-hg} and finding critical vulnerabilities.
Different approaches are needed to fuzz complex targets that require well-structured data types~\cite{Bastani2017-hi, Godefroid2008-sw, Gros2018-di, Han2019-ef, You2019-uu}.
For example, Fuzzil~\cite{Gros2018-di} is a grammar-based fuzzer for JavaScript engines. 
Recent research has focused on fuzzing targets on different platforms~\cite{fuzzware,sgxfuzz,usbfuzz}. Fuzzware~\cite{fuzzware}, SGXFuzz~\cite{sgxfuzz}, and USBFuzz~\cite{usbfuzz} fuzz their targets in an emulated environment. 
\tool follows a similar fuzzing strategy as it generates valid Solana transactions from randomly mutated input bytes and executes them in an emulated Solana blockchain.

Taint propagation~\cite{You2019-uu, Rawat2017-lj, pata, angora, dowser, Aschermann2019-ha} is widely used by fuzzers to identify which part of an input should be changed.
Dowser~\cite{dowser} applies taints by identifying the input data bytes used in security-relevant operations. 
Vuzzer~\cite{Rawat2017-lj}, Redqueen~\cite{Aschermann2019-ha}, and PATA~\cite{pata} concentrate on guiding the fuzzer to pass barriers, e.g., passing a magic value validation. A similar technique could be incorporated in \tool to further improve the performance.
In this paper, we use Libafl~\cite{libafl} as a fuzzer.
Libafl is a high-performance fuzzer with state-of-the-art fuzzing techniques, e.g., including a \emph{persistent mode} which avoids the bottleneck of forking a new process for each fuzzing iteration.
Furthermore, the extensible design of Libafl enables us to integrate \tool's coverage information component into the fuzzing loop.
Therefore, Libafl is the best available fuzzer for this work.

\paragraph{Smart Contract Fuzzing}
Fuzzing has also been applied to Eth-ereum smart contracts~\cite{harvey,confuzzius,rodler2023efcf, contractfuzzer, Ding2021-qg}. 
Harvey~\cite{harvey} is a coverage-guided fuzzer for smart contracts.
It uses program instrumentation to create a feedback mechanism for input prediction.
In addition to that, it creates transaction sequences to detect smart contract bugs, like reentrancy and integer bugs.
ConFuzzius~\cite{confuzzius} is a hybrid fuzzer.
It leverages data dependency analysis and symbolic taint analysis to solve input constraints with the goal of reaching deeper nested paths, thereby increasing coverage.
EFCF~\cite{rodler2023efcf} is a coverage-guided binary-only fuzzer that tightly integrates with well-known fuzzing frameworks for native and legacy programs.
EFCF transpiles the EVM bytecode of a smart contract to native C++ programs and uses state-of-the-art fuzzing optimization techniques~\cite{aflpp}.

However, all these smart contract fuzzing approaches only cover Ethereum smart contracts. 
Ethereum and Solana vulnerabilities are fundamentally different and require different bug detection mechanisms.
In addition, the aforementioned approaches benefit from Ethereum's more advanced security tool landscape, which Solana lacks due to its immaturity, and transferring these techniques require extensive effort.
As a result, these techniques are neither applicable to detect Solana-specific vulnerabilities, nor do they support Solana programs at all.

\paragraph{Smart Contract Analysis}
In addition to fuzzing, other approaches to detect bugs and securing smart have been proposed~\cite{Schneidewind2020-mj, securify, oyente, evmpatch, sereum, Mossberg2019-xp, Torres2018-je}. 
Securify~\cite{securify} deploys formal verification to detect vulnerabilities in Ethereum smart contracts.
It discovers a variety of vulnerabilities, including reentrancy, race conditions, and timestamp dependency.
Oyente~\cite{oyente} and Manticore~\cite{Mossberg2019-xp} are symbolic execution tools which analyze contracts to discover reentrancy and integer overflow issues. 
Sereum~\cite{sereum} deploys dynamic analysis to detect a variety of reentrancy attack patterns. 
EVMPatch~\cite{evmpatch} instruments the bytecode of smart contracts to enable instant patching of smart contracts.
Osiris~\cite{Torres2018-je} introduces a symbolic execution approach for detecting integer-related bugs in Ethereum smart contracts.
While these approaches are successful in detecting bugs in Ethereum smart contracts, they are heavily reliant on Solidity or EVM bytecode, which are closely tied to the Ethereum ecosystem.
Furthermore, many works focus on bug classes that do not exist in Solana, like reentrancy.
To conclude, these approaches cannot analyze Solana programs, and also lack detection capabilities for Solana-specific bugs. 

%!TEX root = ../main.tex

\section{Conclusion}
% Summary
In this paper, we propose \tool, the first coverage-guided fuzzer for Solana programs.
\tool implements five oracles to detect common vulnerabilities in Solana programs, namely missing signer checks, missing owner checks, missing key checks, arbitrary CPIs, and integer bugs.
Moreover, \tool also implements a generic oracle based on lamport gains.
To evaluate the capability of \tool to detect these vulnerabilities, we evaluated \tool against known vulnerabilities~\cite{Neodyme2021-vw}, including the infamous Wormhole bug~\cite{Goodin2022-wormhole}, and compared its performance with VRust~\cite{Cui2022-nm}.
The evaluation has shown that \tool is able to detect the above-mentioned Solana program vulnerabilities and outperforms the state-of-the-art Solana bug detection tool.
We also performed an extensive evaluation in which we fuzzed \noc programs from the Solana mainnet blockchain.
We showed that \tool is the only existing approach to effectively and precisely detect bugs in Solana smart contracts, without relying on source code.
\tool reports 92 bugs in 52 programs, 14 of which we have verified as exploitable bugs at the time of writing the paper.
Moreover, we evaluated \tool's performance against a set of bug bounty programs consisting of complex and productively used Solana programs.
The evaluation has shown that \tool is capable to fuzz complex programs with up to \num{5211} transactions per second.
Our large-scale evaluation is the most extensive security analysis of the Solana blockchain known to date.

%!TEX root = ../main.tex

\section*{Acknowledgment}
Part of this research was conducted within a student project group at the University of Duisburg-Essen.
We thank the project participants, Michael Mboni and Yelle Lieder, for their contribution.
This work has been partially funded by the Deutsche Forschungsgemeinschaft (DFG, German Research Foundation)---SFB 1119 (CROSSING) 236615297 within project T1, EXC 2092 (CASA) 39078197---and the European Union (ERC, CONSEC, No. 101042266, and Horizon 2020 R\&I, DYNABIC, No. 101070455).
The views and opinions expressed are those of the authors only and do not necessarily reflect those of the European Union or the European Research Council Executive Agency.
Neither the European Union nor the granting authority can be held responsible for them.

{
    \printbibliography
}

\end{document}